*Electrically Tunable Heliconical Smectic Superstructure in Polar Fluids*

Hiroya Nishikawa*[,1], Dennis Kwaria[1], Atsuko Nihonyanagi[1], Koki Sano[2], Hiroyuki Yoshida[3] and Fumito Araoka*[,1]

*[1] RIKEN Center for Emergent Matter Science, 2-1 Hirosawa, Wako, Saitama 351-0198, Japan*
*[2] Department of Chemistry and Materials, Faculty of Textile Science and Technology, Shinshu University, 3-15-1 Tokida, Ueda, Nagano 386-8567, Japan*
*[3] School of Engineering, Kwansei Gakuin University, 1 Gakuen Uegahara, Sanda, Hyogo 669-1330, Japan*

E-mail: hiroya.nishikawa@riken.jp; fumito.araoka@riken.jp

**Abstract**

The ferroelectric nematic ($N_F$) phase and related polar liquid-crystalline phases form a new class of strongly polarized yet fluid soft matter. Well-recognized heliconical ferroelectric phases such as $N_{TBF}$ and $SmC_P^H$ exist, but their electro-optic functionality in smectic systems has been largely unexplored. Here, we report a newly designed single-component achiral molecule exhibiting a hierarchical polar phase sequence: $SmA_F$–$N_{TBF}$–HEC–$SmC_P^H$. The key feature of the $SmC_P^H$ phase is its ability to form a stable macroscopic orientation without any alignment layers and to enable continuous, reversible pitch modulation over a wide spectral range at ultralow electric fields, approximately one-third of those required for conventional heliconical nematics. The system exhibits characteristic electro-optic properties in the sub-kilohertz range, differentiating it from heliconical materials controlled by dielectric–elastic balance at higher frequencies. In addition, the $SmC_P^H$ phase's polar heliconical smectic structure facilitates enhanced second-harmonic generation via favorable phase matching, thereby establishing a simple and robust platform for low-voltage photonic applications.

## 1. Introduction

Since its discovery in 2017,[1–3] the ferroelectric nematic ($N_F$) liquid crystal (LC) has been established as a distinct state of fluid matter, where ferroelectric order coexists with high fluidity. In the $N_F$ phase, the head–tail symmetry of the nematic director ($\mathbf{n} = -\mathbf{n}$) present in a conventional uniaxial nematic (N, **Figure 1**a) phase is broken, resulting in a polar structure characterized by aligned director and polarization vectors ($\mathbf{n} \neq -\mathbf{n}, \mathbf{n} \parallel \mathbf{P}$), where $\mathbf{P}$ denotes the polarization vector (Figure 1b).[3–5] This symmetry breaking gives rise to exceptionally high polarization densities and strong nonlinear optical responses while preserving fluidity, thereby underpinning a growing family of polar ferroelectric and helielectric LC phases.[4,5]

Following the discovery of the $N_F$ phase, various related polar LC phases have been identified, introducing hierarchical symmetry breaking into polar fluids through the successive emergence of additional chirality and/or smectic positional order (Figure 1a–e). When smectic layering develops from the $N_F$ phase, a ferroelectric smectic A ($SmA_F$) phase appears (Figure 1c), representing a polar analog of the conventional orthogonal smectic A (SmA) phase with $\mathbf{n} \parallel \mathbf{P}$.[6–8] Further structural complexity arises when molecular tilt develops within the smectic layers, leading to the ferroelectric smectic C ($SmC_F$) phase with $\mathbf{n} \parallel \mathbf{P}$.[9,8,10]

In parallel, the introduction of chirality gives rise to helicoidal superstructures. The chiral nematic (N*) phase is a well-established, nonpolar LC state formed by induced chirality from intrinsically chiral mesogens or chiral dopants.[11] In contrast, the chiral $N_F$ ($N_F$*) phase represents the polar counterpart of N*, where polar helicoidal order arises from the coupling between polar order and chirality. Similar to the N* phase, this phase can be realized either by designing intrinsically chiral polar molecules [12,13] or by introducing chiral dopants [14,15] into the $N_F$ phase. In the $N_F$* phase, inversion symmetry is broken, and the polar structure is inherently chiral.

In 2024, two independent studies demonstrated that achiral molecular systems can spontaneously break chiral symmetry, giving rise to heliconical ferroelectric phases in the absence of molecular chirality. One manifestation of this phenomenon is the heliconical ferroelectric nematic phase ($N_{TBF}$),[16,7] which represents a striking example of spontaneous chiral symmetry breaking within the $N_F$ phase. This concept was further extended to smectic systems with the discovery of the polar heliconical smectic C phase ($SmC_P^H$), which combines long-range smectic layering, molecular tilt, and heliconical polar order.[17,18] Together with the $SmA_F$ and $SmC_F$ phases reported in 2022 and 2024, these discoveries establish a hierarchical symmetry-breaking framework in polar fluids. Along this progression, an intermediate polar heliconical phase—referred to as the helielectric conical (HEC) mesophase—has also been reported.[19] This phase is defined as a polar state with a heliconical structure exhibiting SmC-like local lamellar ordering (Note 1, Supporting Information).

From an application perspective, helical polar phases with pitches in the visible light wavelength range have attracted growing attention as photonic materials. For the $N_F$* and $N_{TBF}$ phases, electric-field-induced [20] and photo-induced [21] color modulation, as well as laser oscillation,[22,23] have been demonstrated. In contrast, previous studies on the $SmC_P^H$ phase have primarily focused on electric-field-induced reorientation of the helical axis,[18] whereas systematic investigations of electric-field-driven modulation of the heliconical pitch itself remain scarce. Consequently, continuous pitch tunability, reflective color modulation, and nonlinear optical functionality in the $\mathrm{SmC_P^H}$ phase remain largely unexplored.

Therefore, the central motivation of this work is to address the following question: can polar heliconical phases serve as electrically tunable photonic media beyond currently reported electro-optic responses? To this end, we build on a previously reported polar molecule (Model I, **Figure 2**a) exhibiting the HEC phase[19] and design a new single-component achiral molecule (**Compound 1**, Figure 2b; see Figures S1–S4, Supporting Information, for characterization spectra) through targeted molecular modification, involving lateral fluorination. The resulting material exhibits a rich phase sequence upon cooling (N–$\mathrm{N_X}$–$\mathrm{SmA_F}$–$\mathrm{N_{TBF}}$–HEC–$\mathrm{SmC_P^H}$), reflecting hierarchical symmetry breaking. Notably, the $\mathrm{SmC_P^H}$ phase in **Compound 1** forms a stable macroscopic orientation under an applied electric-field (*E*-field) without alignment layers. More importantly, it enables continuous *E*-field-driven heliconical pitch modulation at ultralow voltages and in the subkilohertz frequency regime, accompanied by vivid reflective color control and enhanced second-harmonic generation (SHG). To the best of our knowledge, such continuous and reversible pitch modulation with visible reflective color tuning has not been reported for previously known $\mathrm{SmC_P^H}$ materials. These combined features distinguish the present material from previously reported heliconical systems and reveal a dynamically mediated route to electrically tunable structural color in heliconical smectic systems.

## 2. Results and Discussion

### 2.1. Phase transition behavior

The differential scanning calorimetry (DSC) curve (Figure 2c; Figure S5 and Table S1, Supporting Information) and polarized optical microscopy (POM) images of **Compound 1** (Figure 2d–s; Figure S6, Supporting Information) were used to characterize its phase transition behavior. For POM studies, we used syn-parallel rubbing polyimide (SP) glass cells (Figure 2d–h), bare substrate cells (Figure 2i–m), and indium tin oxide (ITO)-coated glass cells (Figure 2n–s). In the SP cell, cooling from the nematic (N) phase caused many defects (banded texture) to appear along the rubbing direction (Figure S6b, Supporting Information), which then developed into uniform block-like domains aligned in the rubbing direction (Figure 2d). These features are typical in the $\mathrm{N_F}$ phase transiting into the $\mathrm{SmA_F}$ phase. However, this phase (hereinafter refer to as the $\mathrm{N_X}$ phase) exhibited a high apparent relative dielectric permittivity ($\varepsilon'_{app}$ = 5.2 × $10^3$) (*vide infra*) despite showing an antiferroelectric-like electrical response (Figure S16, Supporting Information), suggesting antiferroelectric, ferrielectric or relaxor-like[24] behavior. Due to it extremely narrow temperature range, no further classification is performed. Further cooling revealed a stripe texture along the rubbed direction, with additional finer stripes appearing perpendicular to the initial stripes (Figure 2e). Within this temperature range, colored reflection or scattering was visible in the LC cell. This was more obvious in a bare substrate cell (Figure 2j). The full pitch of the fine stripes corresponds to that of the heliconical structure, which was estimated to be 1.3 μm (Figure S7, Supporting Information). Typical POM textures reported for the $N_{TBF}$ phase were observed in the present study.[16, 7] In the bare-substrate cell, the appearance of a vivid selective-reflection color is consistent with a heliconical structure whose helical axis tends to align nearly normal to the substrates, thereby enhancing selective reflection. In the SP cell, stripe textures aligned along the rubbing direction are observed, consistent with a heliconical structure with the helical axis lying in-plane. These characteristic textures are in good agreement with our experimental observations and support the assignment to the $N_{TBF}$ phase. Note that this assignment is further supported by the combined evidence from X-ray diffraction (XRD), broadband dielectric spectroscopy (BDS), polarization reversal current (PRC) and SHG measurements presented in the subsequent sections. It is noteworthy to mention that Strachan et al. discussed the possible competition of the $SmA_F$ and $N_{TBF}$ phases [25], and that no $N_{TBF}$–$SmA_F$ phase transition within a single component molecule has yet been reported to date. The $N_X$–$SmA_F$–$N_{TBF}$ phase transition in **Compound 1** is obviously one of the rare examples of the reentrant phenomenon in polar phases [26], where the restoration of nematic order follows the collapse of smectic layer order, although the mechanism of this is not yet clear. The $N_{TBF}$ phase existed only within a narrow temperature range of approximately 1 K, after which inhomogeneous textures of the HEC phase appeared (Figure 2e,f; Figure S8a–c, Supporting Information). Such a textural change was also observed in thicker/thinner bare substrate (Figure 2j–l; Figure S8d–f, Supporting Information) and thinner ITO cells (Figure S8g–i, Supporting Information). Interestingly, in the bare substrate cell, a coil-like spiral filament texture was observed in the same temperature region (Figure 2k and 2l). The similar spiraling filaments were also observed in the HEC phase in our previous study.[19] In the SP cell, the faint reflective color was red-shifted upon cooling, and eventually the texture transited into a darker state at 84℃ (Figure 2g, 2h; Figure S6f–l, Supporting Information). A texture with low birefringence was also observed at 85 ℃ in the bare substrate cell (Figure 2m). The ITO-coated cell within this temperature range displayed sub-millimeter-sized island domains in which the contrast between adjacent domains was reversed by decrossing the polarizers in the opposite directions, indicating optical activity (Figure 2n–p). Furthermore, reflection microscopy confirmed reflective colors (Figure 2q–s) of left- or right-handed circularly polarized selective reflection. Given the generation of enantiomeric domains and the observation of selective reflective color, the low-temperature phase appearing below the HEC phase was considered to be the $\mathrm{SmC_P^H}$ phase. So far, the control over both the reflective color and the orientation of the helical axis in the $\mathrm{SmC_P^H}$ phase is yet difficult [17,18]. In contrast, the $\mathrm{SmC_P^H}$ phase in **Compound 1** enables both *E*-field-induced reorientation of the helical axis and color modulation via the helical pitch variation, as will be demonstrated in the below.

**2.2. Structural characterization**

Two-dimensional (2D) wide-angle XRD patterns were obtained for structural characterization of **Compound 1** at various temperatures (**Figure 3**a–d; Figures S9 and S10, Supporting Information). The corresponding 1D XRD plots were generated by the box scan analysis (Figure S11, Supporting Information) along the equator parallel to the applied magnetic field **B** (**n** ∥ **B**) of the 2D XRD pattern (Figure 3e, 3f; Figure S12, Supporting Information). The following parameters were then obtained as functions of temperature: *d*-spacing ($d = 2\pi/q$) (Figure 3g), tilt angles ($\theta_{\mathrm{tilt}}$) (Figure 3g), full width at half maximum (FWHM) of the scattering vector, $q$ ($\Delta q_{\mathrm{FWHM}}$) (Figure 3h), and the peak intensity (Figure 3h). The diffraction pattern of the N phase can be attributed to short-range positional ordering of molecules along **B** (Figure S10a, Supporting Information). Although the peak intensities and positions were almost the same in the N and $\mathrm{N_X}$ phases (Figure S10b, Supporting Information), $\Delta q_{\mathrm{FWHM}}$ was smaller in the $\mathrm{N_X}$ phase than in the N phase. After the transition to the $\mathrm{SmA_F}$ phase, the diffraction peak became sharper, and $\Delta q_{\mathrm{FWHM}}$ was greatly reduced, suggesting an increase in the correlation length. In the $\mathrm{SmA_F}$ phase, *d* became almost consistent with the molecular length (*L*) due to the formation of the non-tilted lamellar structure. After transition to the $\mathrm{N_{TBF}}$ and HEC phases, the diffraction pattern became a diffuse arc. In such a heliconical structure, the local director is tilted by a certain angle with respect to the helical axis and continuously rotates around it. As a result, even when the average orientation of the helical axis is aligned, the local orientations are distributed over a conical surface in real space. This angular distribution leads to diffuse arc-shaped scattering in reciprocal space rather than a sharp diffraction spot. Within the HEC phase temperature range, *d* gradually decreased upon cooling, meaning that the molecules in the lamellar were tilted with respect to the layer normal. The molecular tilting started at the lower-temperature side of the $\mathrm{SmA_F}$ phase, with the tilt angle measured as $2° < \theta_{\mathrm{tilt}} < 7°$ where $\theta_{\mathrm{tilt}} = \arccos\left(\frac{d}{L}\right)$. Peak intensity, *d* and $\Delta q_{\mathrm{FWHM}}$ jump at the HEC–$\mathrm{SmC_P^H}$ phase transition, suggesting the first-order type. In the $SmC_P^H$ phase, the molecular orientation with respect to the magnetic field deviated substantially.The $SmC_P^H$ phase exhibited further gradual reduction in *d*, leading to a maximum tilt angle $\theta_{tilt}$ = 21°, despite an almost constant $\Delta q_{FWHM}$. On the other hand, in the HEC phase, $\Delta q_{FWHM}$ increased and then reached a plateau. This suggests that, although both the HEC and $SmC_P^H$ phases possess polar heliconical ordering of the tilted smectics, the coherence of the structure in the HEC phase was reduced upon cooling, implying the generation of small fragments or clusters of the smectic ordering.

### 2.3. Polar behavior

Next, the polar behavior of various phases was evaluated using BDS, PRC and SHG measurements.

As shown in **Figure 4**a–c and Figures S13 and S14 (Supporting Information), the BDS results revealed that $\varepsilon'_{app}$ [27,28] increased to 5.2 × $10^3$ at the N–$N_X$ phase transition point, decreased sharply to a minimum of 444 in the $SmA_F$ phase, and recovered to $\varepsilon'_{app}$ of 11 × $10^3$ in the $N_{TBF}$ phase. This sequence of permittivity changes is likely attributed to the suppression of polarization fluctuations [8] and the restoration of a strong collective relaxation mode [29,16] in the $SmA_F$ and $N_{TBF}$ phases, respectively (Figure 4b). In the HEC phase, $\varepsilon'_{app}$ remained 10 × $10^3$ in the high-temperature region but gradually decreased toward the low-temperature region, reaching a plateau at approximately 2.7 × $10^3$. Upon the phase transition from the HEC phase to the $SmC_P^H$ phase, the dielectric permittivity recovered to 7.6 × $10^3$ and then gradually decreased to 1.7 × $10^3$. As shown in Figure 4b, the $N_F$ and $N_{TBF}$ phases exhibited a dominant low-frequency dielectric relaxation mode, which can be attributed to the collective polar fluctuations.[28,5,16] In the HEC phase, this low-frequency mode persisted, while additional relaxation featured emerge at higher frequencies upon cooling, indicating the presence of multiple relaxation processes. This behavior was agreement with the previous results for the HEC phase.[19] As confirmed by XRD measurements, the $SmC_P^H$ phase exhibited a tilted smectic ordering, which was consistent with a Goldstone mode associated with the azimuthal rotation on the tilt cone[30] (Figure 4c). In contrast, the HEC phase showed smectic-like short-range correlations, which may be interpreted as cybotactic-like local ordering, with no evidence for long-range smectic order.[19]

PRC measurements in the $SmA_F$, $N_{TBF}$, HEC, and $SmC_P^H$ phases (Figure 4d–g; Figures S15 and S16, Supporting Information) were carefully conducted upon checking the phase transitions by POM. Notably, the phase transition temperatures inevitably differed between BDS and PRC measurements because of the different cell conditions, such as thickness, used in each experiment. The temperature dependence of spontaneous polarization ($P_s$) was determined using the PRC results (Figure 4h). The $N_X$ phase exhibited $P_s$ = 1.6 μC cm$^{-2}$. In the $SmA_F$ phase, $P_s$ depended on temperature, ranging between 1.9 and 3.0 μC cm$^{-2}$, whereas the $N_{TBF}$ phase exhibited an almost constant $P_s$ = 3.0 μC cm$^{-2}$. Two distinct current peaks were observed in the entire HEC region—one at the crossing point of the *E*-field at zero and the other slightly after that. The *E*-field at low frequency (20 Hz) clearly split the current signal into two peaks, suggesting that the first and second peaks correspond to relaxation and spontaneous polarization reversal, respectively (Figure S17, Supporting Information). Thus, the peak splitting does not imply multiple polarization states. In the $\mathrm{SmC_P^H}$ phase, the relaxation contribution is suppressed due to the development of smectic ordering, resulting in a single current peak and a continuous increase of $P_s$. Within the $\mathrm{SmC_P^H}$ region, $P_s$ increases gradually with decreasing temperature, reaching values of 4.6 μC cm$^{-2}$ within the investigated temperature range; this large $P_s$ reflects a well-developed smectic ordering in the $\mathrm{SmC_P^H}$ phase. In addition, we confirmed that the $\mathrm{SmC_P^H}$ phase exhibited high cycle stability over 1000 switching cycles (Figure S18, Supporting Information).

Finally, temperature-dependent SHG measurements were performed (Figure 4i). Strong SHG signals were detected in all polar phases except the $\mathrm{SmC_P^H}$ region, where SHG signal was strongly suppressed but remained finite. In the $\mathrm{SmC_P^H}$ region, the sample area in the glass cell became cloudy, likely due to scattering mode, in which the helical axis randomly orients, such as cholesteric LCs. Consequently, the incident beam was scattered, potentially preventing accurate SHG measurements. More details on SHG are provided in Section 2.5.

### 2.4. Color modulation of heliconical smectic superstructure

The $\mathrm{SmC_P^H}$ phase initially exhibited a strongly turbid state, yet it demonstrated the faint selective reflection color due to scattering. This turbidity may have been caused by the random orientation of the helical axes. Similarly to the heliconical phases including the $N_{TBF}$ phase, the molecules in the $\mathrm{SmC_P^H}$ phase were tilted at an angle $0 < \theta < \pi/2$ with respect to the helical axis (= layer normal) (Figure 1d, e). Since in the previous heliconical phases, applying the *E*-field along the helical axis modulates the helical pitch ($p$) and reflective color according to the continuous change in $\theta$, without reorienting the helical axis (**Figure 5**a). Thus, it is natural to expect the field tunable modulation of the reflected color through the pitch modulation also in the present $\mathrm{SmC_P^H}$ phase.

Here, an out-of-plane *E*-field was applied to the cell to examine the influence of the *E*-field on the $\mathrm{SmC_P^H}$ structure. The application of either DC or rectangular AC *E*-field (100 Hz) perpendicular to the electrodes deformed the initial multi domains. When the *E*-field was applied, the helical axis gradually tilted upward along the direction of *E*-field, concurrently with the progressive formation of enantiomeric domains. (Figure S19, Supporting Information). And then, by gently reducing the *E*-field strength, large domains grew and eventually exhibited vivid red reflection at 0 V. We refer this process to as the *pretreatment process*. Although the texture still showed slight nonuniformity and light scattering even after this process, a vivid selective reflection originating from Bragg reflection of the heliconical structure was observed, accompanied by a minimal birefringent background. This state is attributed to a homeotropically aligned $\mathrm{SmC_P^H}$, in which the heliconical axis and the layer normal were parallel to the applied *E*-field. After this pretreatment, when an AC field was applied, the reflected color shifted from red to orange, yellow, green, and blue as gradually increasing the AC *E*-field and vice versa as decreasing the AC *E*-field (Figure 5c–h). When these domains were observed through a left- or right-handed circular polarizer (LCP or RCP), a clear and reversible contrast of bright and dark domains was observed, confirming the chirality in each domain (Figure 5c–h; Figure S20, Supporting Information). Interestingly, the boundary regions between the domains with opposite chirality showed almost identical reflections both for LCP and RCP, while the crossed-Nicol images exhibited extremely low transmittance (Figure S21, Supporting Information). This operates on the same principle of hyper-reflective cholesteric LCs [31,32], which indicates that the left- and right-handed structures were superimposed in this region. Note that such an overlapped region was not observed in thinner cells (4.5 μm) (Figure S22, Supporting Information). In the case of the DC *E*-field, the reflected color changed only slightly from red to reddish orange even under a sufficiently high *E*-field. The difference between DC and AC driving may partially arise from interfacial charge redistribution under static DC bias (Note 2, Figures S23 and S24, Supporting Information), which can reduce the effective internal *E*-field and thereby limit the achievable pitch modulation.

Further, we investigated the spectroscopic properties and their response to the rectangular AC *E*-field at a frequency of 100 Hz (Figure 5i; Video S1, Supporting Information). First, the transmission spectrum showed a reflection band in the visible light to the near infrared spectral range (Figure S25, Supporting Information). The optical response is determined by the half pitch of the polar heliconical structure, as it is governed by the modulation of the refractive index. As a result, the observed reflection band corresponds to the half-pitch Bragg reflection. When the *E*-field was increased, the helical pitch decreased according to the decrease of the tilt angle, which caused blueshift of the reflection band. This electrically tunable reflection is reversible with almost no hysteresis between the forward and backward processes (Figure 5i and 5k; Figure S26a; Video S2, Supporting Information). The required *E*-field to induce pitch modulation—and thereby tune the reflected color among red, green, and blue—was within the range of 0.46–0.77 $V_{pp}$ $\mu m^{-1}$ (0.23–0.39 $V_{rms}$ $\mu m^{-1}$), which was compared with conventional heliconical nematic-type systems [33–39,16,7] as summarized in Figure S27 (Supporting Information). The present heliconical smectic system ($SmC_P^H$) achieves comparable pitch tuning under substantially lower *E*-fields, with the exception of the system reported in Ref. 7. Specifically, the *E*-field required to reach the red region is 2.1–6.2 times lower than that reported for heliconical nematic-type systems, while reductions of 1.9–4.3 times and 1.9–4.1 times are observed for the green and blue regions, respectively. Furthermore, we examined the shift of the reflection band upon varying AC frequency while keeping the strength constant. When the frequency was increased from 20 Hz to 400 Hz at $E$ = 0.63 $V_{pp}$ $\mu m^{-1}$, the selective reflection band redshifted by approximately 150 nm. Then, the shift became saturated at higher frequencies (Figure 5j, l; Video S3, Supporting Information). Conversely, decreasing the frequency resulted in blueshift without hysteresis (Figure S26b, Supporting Information). Note that the shift showed a linear dependence on the frequency between 20 and 400 Hz in both ascending and descending processes. Thus, the tunable color modulation can be achieved even by changing the frequency (Figure S28; Video S4, Supporting Information). In such a way, we demonstrated the straightforward modulation of the helical pitch based on both the strength and frequency of the *E*-field, which enabled color modulation across the visible light spectral region (Figure 5b). Although the above pitch modulation was optically stable on a timescale longer than the period of the applied AC field, certain fluctuations in color and brightness were observed in the texture under POM. The pattern of the fluctuations propagated radially, or sometimes spirally, or in more complex manners, synchronously with the applied AC *E*-field (Video S5 and S6, Supporting Information), so that we consider that this is attributed to the director rotation around the smectic cone. This dynamic behavior looks to be highly intriguing as a new nonequilibrium phenomenon in the polar fluid system.

The optical responses in the present system are more complex than those in conventional heliconical nematic systems, where the competition between elastic and electric-field-induced torques results in a linear dependence of the reflection wavelength on the reciprocal *E*-field strength.[34] As shown in Figure S29 (Supporting Information), the similar $\lambda \propto 1/E$ scaling is observed in the low-field regime (I). However, the intermediate-field regime (II) exhibits a nonlinear response, while the high-field regime (III) again shows a linear dependence. Such complex behavior suggests the interplay of multiple competing factors, including torques arising from dielectric and polar couplings,[40] the elastic restoring force, and hydrodynamic effects. Although a complete description of the mechanism remains challenging, this continuous pitch change is highly reproducible, thereby ensuring reliable control of the pitch modulation.

The frequency dependence further highlights the fundamentally different driving mechanism in the present system. At a fixed field strength, decreasing the frequency induces blue shift, whereas increasing the frequency leads to red shift, with saturation occurring above approximately 400 Hz (Figure 5j,l). This behavior cannot be explained by simple ionic screening. In general, ionic charge accumulation at the electrodes reduces the effective internal field more strongly at low frequency, which would be expected to suppress pitch modulation and induce red shift. Therefore, the observed blue shift at low frequencies rules out a simple ionic-screening origin for the AC-frequency dependence. In contrast, under static DC bias, interfacial charge redistribution can gradually develop and partially screen the effective internal field acting on the $\mathrm{SmC_P^H}$ heliconical structure. Such static screening can limit further pitch modulation under DC driving, whereas AC driving sustains dynamic polarization modulation without forming a persistent static screening state. Therefore, the possible charge-screening effect under DC bias does not contradict the above discussion regarding the AC-frequency dependence.

### 2.5. SHG enhancement on a heliconical smectic superstructure scaffold

In the conventional nonlinear optics, the efficiency of SHG is enhanced by phase matching between the fundamental and harmonic waves.[41] In ferroelectric crystals, quasi-phase matching is enabled by patterning an alternating polarization structure, leading to obtain efficient SHG.[34] In recent years, SHG enhancement based on the nonclassical phase matching has been successively reported in the $\mathrm{N_F}^*$ phase, where the phase matching condition based on photonic bandgap resonance is effectively satisfied, yielding significant SHG amplification[12,42]. Even in the $\mathrm{SmC_P^H}$ phase, the helicoidal polarization structure with the $C_\infty$ symmetry may provide the similar optical condition, leading to the nonclassical phase matching to enhance the SHG efficiency (**Figure 6**a).

We propagated a polarized fundamental beam (Nd:YAG, 1064 nm) along the helical axis normal to the substrate to investigate the SHG response under the helical pitch modulation induced by *E*-field. As expected, the SHG signal was enhanced at $E = 0.59\ \mathrm{V_{pp}}\ \mu\mathrm{m}^{-1}$, and more interestingly, significantly amplified at $E = 0.74\ \mathrm{V_{pp}}\ \mu\mathrm{m}^{-1}$ (Figure 6b). By comparing the reflectance spectra at these voltages, we found that the short-wavelength skirt of the former and the long-wavelength skirt of the latter overlap around the SHG wavelength (532 nm). This suggests that the SHG enhancement originates from phase matching[43] coupled with the short-and long-wavelength band-edge modes at $E = 0.74$ (band (ii) in Figure 6b) and 0.59 $V_{pp}$ $\mu m^{-1}$ (band (i) in Figure 6b), respectively. It is well known that, for light propagation in the helical structure of the conventional N* phase, the long- and short-wavelength band edges correspond to the extraordinary and ordinary propagation modes, respectively. In the $SmC_P^H$ case, the nonlinear polarization in the *x–y* plane is nearly parallel to the *c*-director and can therefore couple with the extraordinary mode, leading to a more pronounced SHG enhancement (Figure 6a). Even then, enhancement also occurs weakly at the short-wavelength band edge, corresponding to the ordinary mode, due to the contribution from the slight off-axis component of the nonlinear polarization. Such band-edge enhancement was also predicted by optical simulation[44]. This nonclassical phase-matching behavior in the $SmC_P^H$ phase provides distinct and tunable SHG enhancement that can be achieved using a low *E*-field applied along the helical axis, unlike in the $N_F^*$ phase.[12,42] These findings offer a design guideline for tunable and efficient nonlinear optical elements.

**2.6. Comparison between the conventional heliconical nematic and the present polar heliconical smectic systems.**

In the final section, we contrast heliconical nematic systems with the polar heliconical smectic system ($SmC_P^H$) proposed in this work. The key properties of these systems are summarized from multiple perspectives in Figure S30 (Supporting Information). Importantly, this comparison shifts the focus from phase classification to material-specific functionality, emphasizing that the key advance lies not in the phase designation itself but in the emergence of a distinct electro-optic response unique to this material.

The most distinctive feature of the present material is its ability to achieve continuous and reversible pitch modulation over a wide spectral range under an ultra-low electric field (approximately one-third of that required for heliconical nematics [39]). Second, this functionality is realized in a single-component system and exhibits excellent reproducibility, in contrast to known heliconical nematic systems that rely on multicomponent mixtures or dopants. Third, the absence of alignment layers suggests that the polar smectic nature of the phase may contribute to stabilizing the macroscopic orientation. In particular, achieving uniform homeotropic alignment remains challenging in the $N_{TBF}$ phase. In addition, the polar nature of the system enables access to even-order nonlinear optical processes. Notably, the helielectric structure provides favorable phase-matching conditions, resulting in enhanced second-harmonic generation, as demonstrated above. Combined with its single-component nature, high optical quality, and straightforward control of the helical axis, this system offers a simple and robust platform for photonic applications, including cavity-less lasing and up-conversion devices. [45]

## 3. Conclusion

In this study, we performed a detailed analysis of the $SmA_F$–$N_{TBF}$–HEC– $SmC_P^H$ phase transitions in a novel polar molecule (**Compound 1**), demonstrating hierarchical symmetry breaking. **Compound 1** was designed through lateral fluorination of a previously reported molecule (Model I), which exhibited the $N_F$–HEC phase transition sequence. The molecular parameters obtained by DFT calculations for Model I and **Compound 1** are summarized in Table S2 (Supporting Information), showing that this single-fluorine substitution modifies the molecular dipole moment and its orientation relative to the long molecular axis. However, the effect of fluorination on polar LC phases should not be interpreted solely in terms of changes in polar properties, because fluorination may also influence local electrostatic interactions, conformations, and molecular packing and ordering. [46–48] Thus, fluorination is considered as one of the sensitive factors responsible for the alteration of the phase sequence ($SmA_F$–$N_{TBF}$–HEC–$SmC_P^H$ in the present case), rather than as the sole or direct origin of the phase behavior. Future theoretical frameworks may further rationalize this structure–phase relationship.[46–49]

During the phase transition from HEC to $SmC_P^H$, a distinct first-order phase transition was observed, accompanied by characteristic changes in POM textures, as well as discontinuous behaviors in BDS, PRC, and SHG, all of which clearly demonstrate the structural transition. Furthermore, $E$-field-driven pitch modulation in the $SmC_P^H$ phase reveals behavior consistent with a polarization-coupled dynamic mode operating in the subkilohertz regime. Whereas inverse scaling of the reflection wavelength with $E$-field is observed in the low- and high-voltage regimes, the intermediate regime exhibits deviations from simple scaling, together with reduced reflectance, indicating nonlinear polarization coupling and increased spatial heterogeneity. The combined observation of scaling behavior, nonlinear deviation, and a characteristic frequency cutoff of 400 Hz distinguishes this system from previously reported heliconical materials governed by a frequency-dependent dielectric–elastic balance at high frequencies.[38, 50] These findings suggest that dynamic polarization governs the electro-optic functionality of the $SmC_P^H$ phase.

Taken together, this work not only advances the understanding of hierarchical polar order in soft matter but also demonstrates a material-specific, low-frequency and ultralow-voltage route to electrically tunable structural color in smectic heliconical systems, offering new design principles for polar soft-matter photonics.

## Data availability

The authors declare that the data supporting the findings of this study are available within the paper and its supporting information files. All other information is available from the corresponding authors upon reasonable request.

## Acknowledgements

We are grateful to Dr. H. Koshino (RIKEN, CSRS) for allowing us to use JNM-ECZ500 (500 MHz, JEOL). We wish to thank Dr. Nogawa (RIKEN, CSRS) for the HRESI-MS measurement. We would like to acknowledge the Hokusai GreatWave Supercomputing Facility (project no. RB230008) at the RIKEN Advanced Center for assistance in computing and communication. This work was partially supported by JSPS KAKENHI (JP22K14594; H.N., 23K26573; H.Y., JP21H01801, JP23K17341; F.A.), RIKEN Incentive Research Projects (FY2024: H.N.), JST CREST (JPMJCR23O1; K.S., F.A.) and JST SICORP EIG CONCERT-Japan (JPMJSC22C3; F.A.).

## Contributions

H.N. and F.A. conceived the project and designed the experiments. H.N. designed molecules and performed DFT calculation. A.N. synthesized and characterized all compound. H.N. carried out DSC, POM, BDS, PRC and UV-vis-NIR spectra studies. H.Y. constructed reflection spectra measurement system and recorded spectra. D.K. performed SHG measurements. K.S. and H.N. took XRD measurements. H.N. and F.A. analyzed data and discussed the results. H.N. and F.A wrote the manuscript, and all authors approved the final manuscript.

## Corresponding authors

Correspondence to Hiroya Nishikawa or Fumito Araoka

## Competing interests

The authors declare no competing interests.

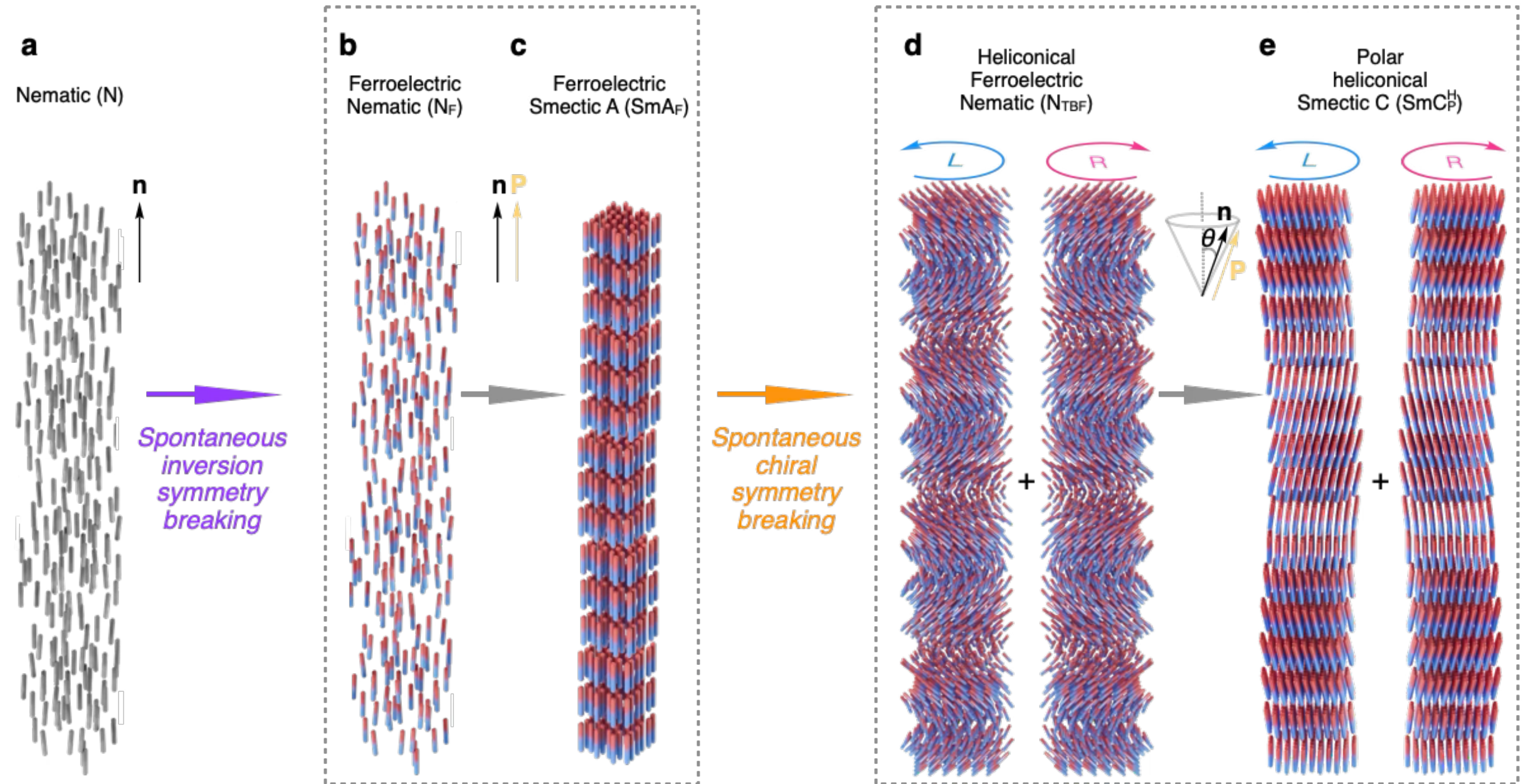

**Figure 1** Schematic of the phase transition sequence for a polar material via hierarchical symmetry breaking (**n**: director and **P**: polarization vector). All phases except for the nematic (N) phase lack a head–tail orientation equivalence, so polar order occurs (i.e., $\mathbf{n} \neq -\mathbf{n}$). The $N_F$ phase emerge via spontaneous inversion symmetry breaking from the isotropic or N phase. The same pathway offers a transition to the $SmA_F$ phase with additional layer positional ordering. Subsequent spontaneous chiral symmetry breaking generates unique helielectric phases (i.e., $N_{TBF}$ and $SmC_P^H$) in which n and P are tilted at an angle $0 < \theta < \pi/2$ with respect to the helical axis. In both phases, left ($L$)- and right ($R$)-handed helices are formed with equal probability.

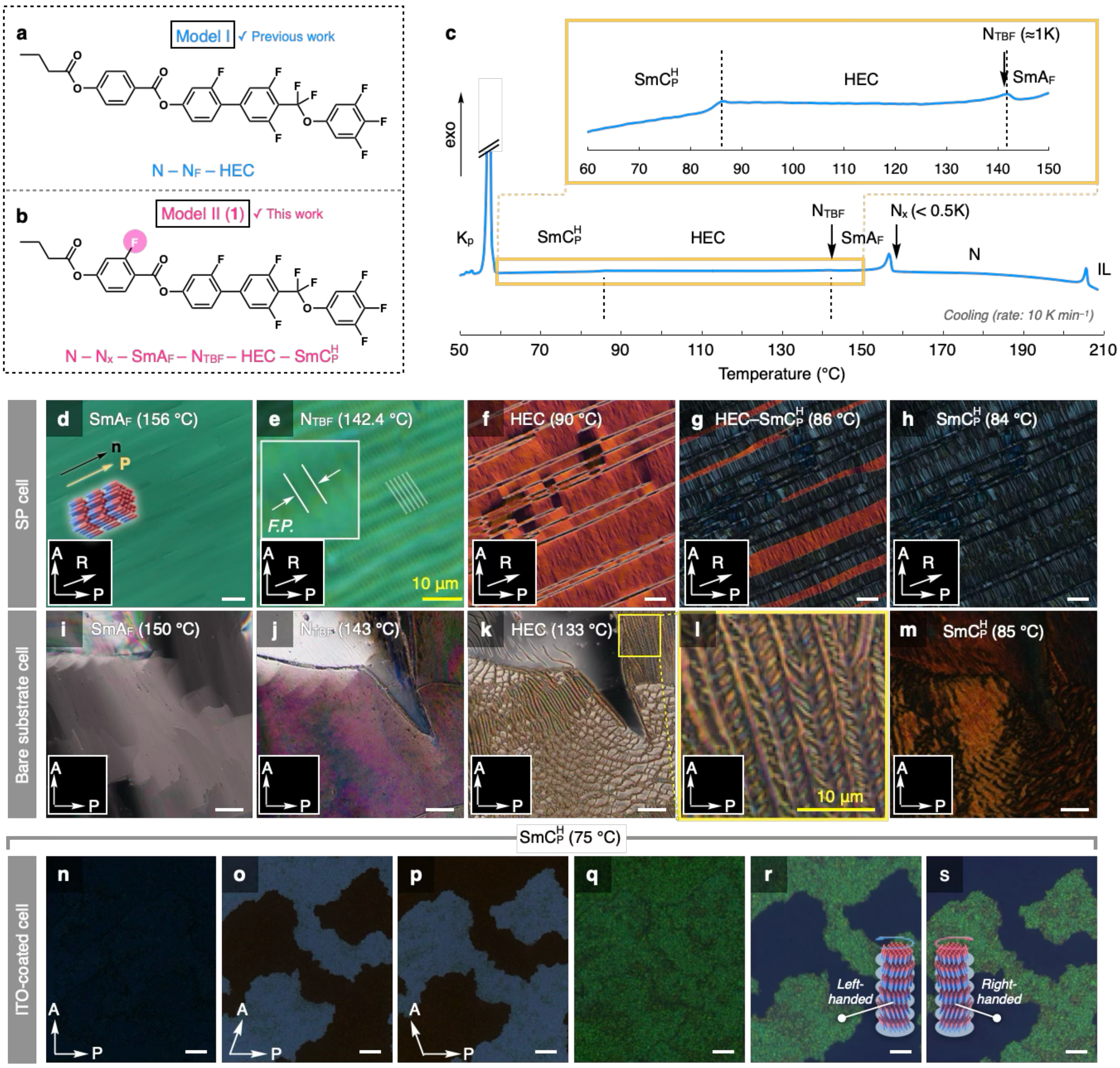

**Figure 2** Phase transition behavior for **Compound 1**. Chemical structures used in the (a) previous study (Model I) and (b) present study (Model II or **Compound 1**). (c) DSC curves during cooling (scan rate: 10 K min$^{-1}$). POM images taken under cross polarizers in SP (d–h), bare substrate (i–m), and ITO-coated glass (n–s) cells at various temperatures. Thickness: SP cell: 5.0 µm; bare substrate cell: 25 µm; ITO cell: 4.6 µm. Reflection-mode POM images of the $SmC_P^H$ phase ($T$ = 75°C) (q) without a circular polarizer and with (r) left-handed and (s) right-handed circular polarizers. The inset in the panel (e) shows the enlarged photograph, in which the distance between dark lines corresponds to the full pitch (F.P.) of $N_{TBF}$ phase. The inset in panel (d) shows a schematic illustration of the $SmA_F$ phase. In the SP cell, smectic blocks lay on the substrate, and molecules were aligned along the rubbing. The insets in panels (r) and (s) show schematic illustrations of left- and right-handed polar heliconical structures. Scale bars: 100 µm (except the panels (e) and (l)).

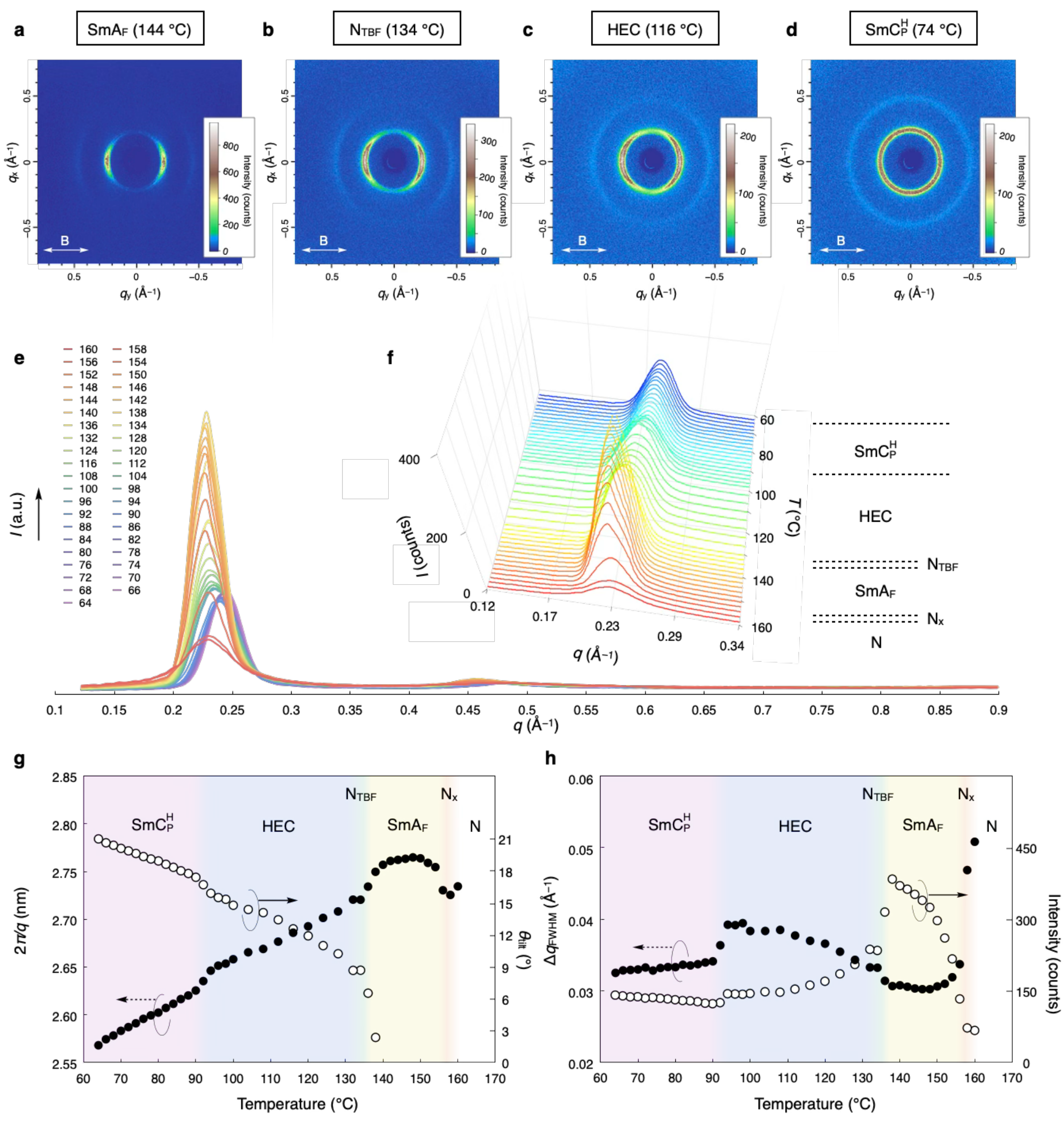

**Figure 3** XRD profiles for **Compound 1**. 2D WAXS patterns obtained under a magnetic field (~1 T, B ∥ n) of the (a) SmA$_F$, (b) N$_{TBF}$, (c) HEC, and (d) SmC$_P^H$ phases. (e) 1D and (f) 3D X-ray diffractograms generated by box-scan analysis of the 2D WAXS profile. (g) $2\pi/q$ and tilt angle ($\theta_{tilt}$) vs temperature. (h) $\Delta q_{FWHM}$ and intensity vs temperature.

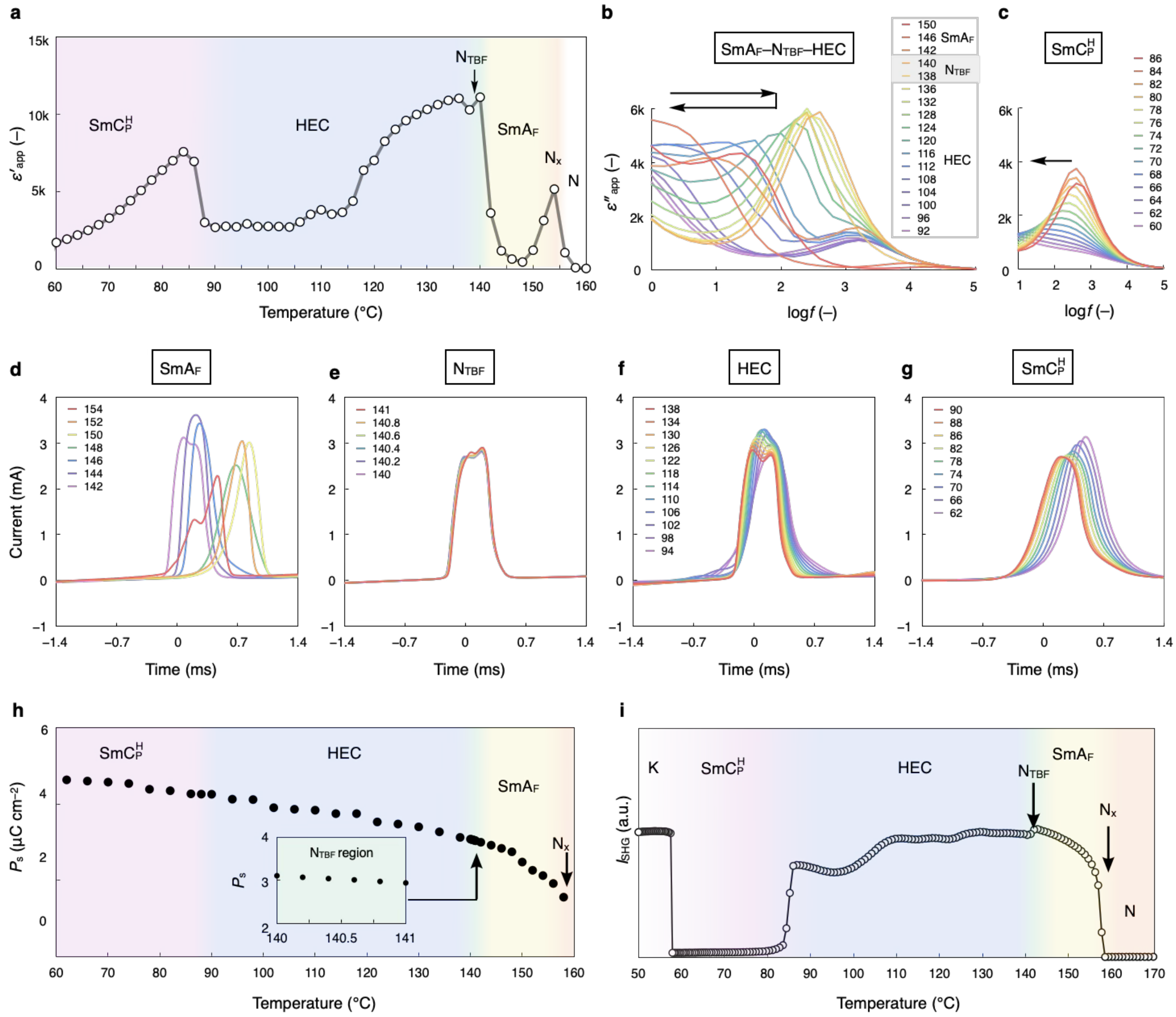

**Figure 4** Polar behavior of **Compound 1**. (a) Apparent relative dielectric permittivity ($\varepsilon'_{app}$) vs temperature. Dielectric loss ($\varepsilon''$) vs frequency for the (b) SmA$_F$–HEC phase transition and (c) in the SmC$_P^H$ phase. Black arrows indicate direction of peak shift. PRC profiles of the (d) SmA$_F$, (e) N$_{TBF}$, (f) HEC, and (g) SmC$_P^H$ phases. Temperature-dependent (h) spontaneous polarization $P_s$, and (i) SHG.

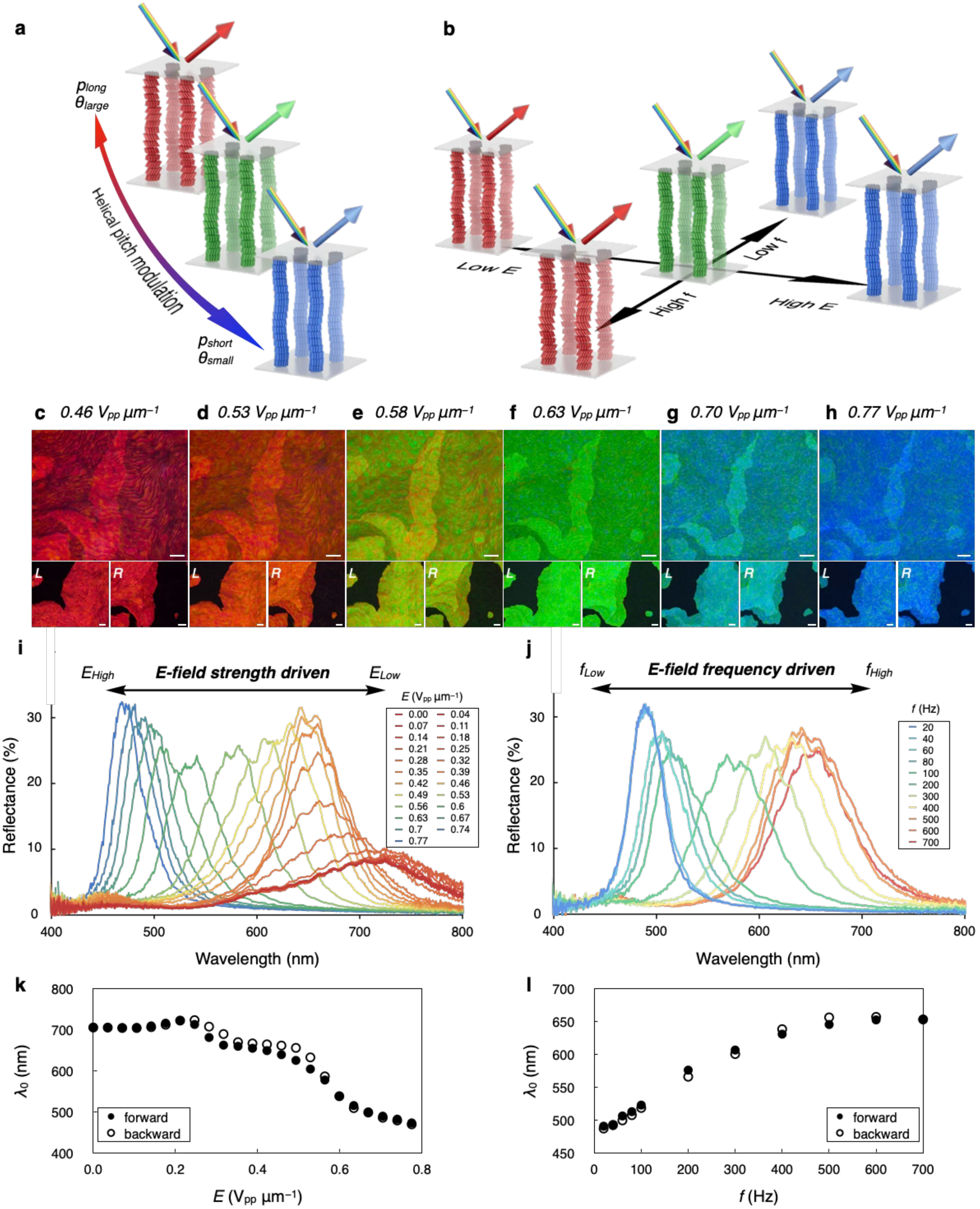

**Figure 5** Color modulation in the SmC$_P^H$ phase of **Compound 1** ($T$ = 75°C). Schematic illustrations of (a) modulation of the reflected color associated with the helical pitch and (b) color modulation driven by the $E$-field strength and frequency. (c–h) Evolution of reflected color in reflection-mode POM (RPOM) images taken after pretreatment process. The bottom images were taken with left- and right-circular polarizers (symbol: L and R). Scale bar: 200 μm. Evolution of reflection spectra by modulation of the (i) $E$-field strength and (j) frequency. Central wavelength ($\lambda_0$) vs (k) $E$-field strength and (l) $E$-field frequency.

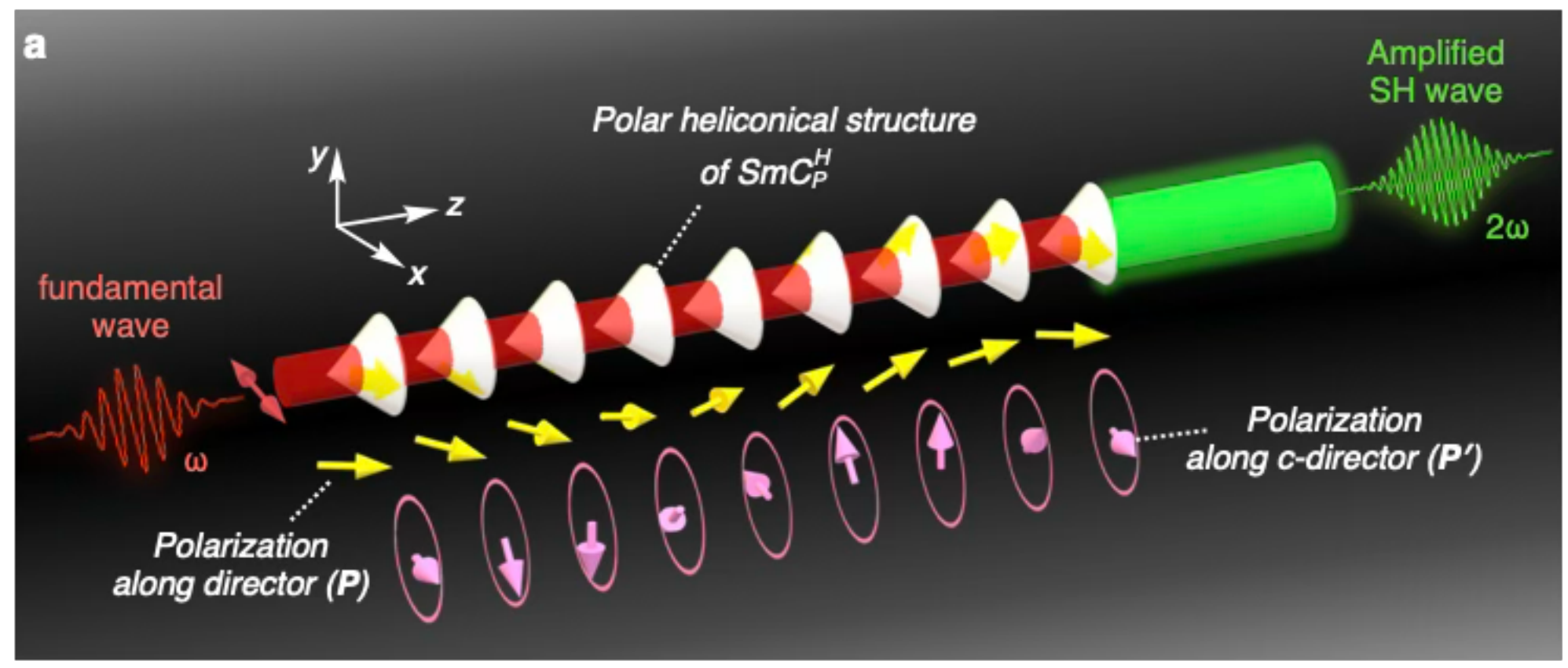

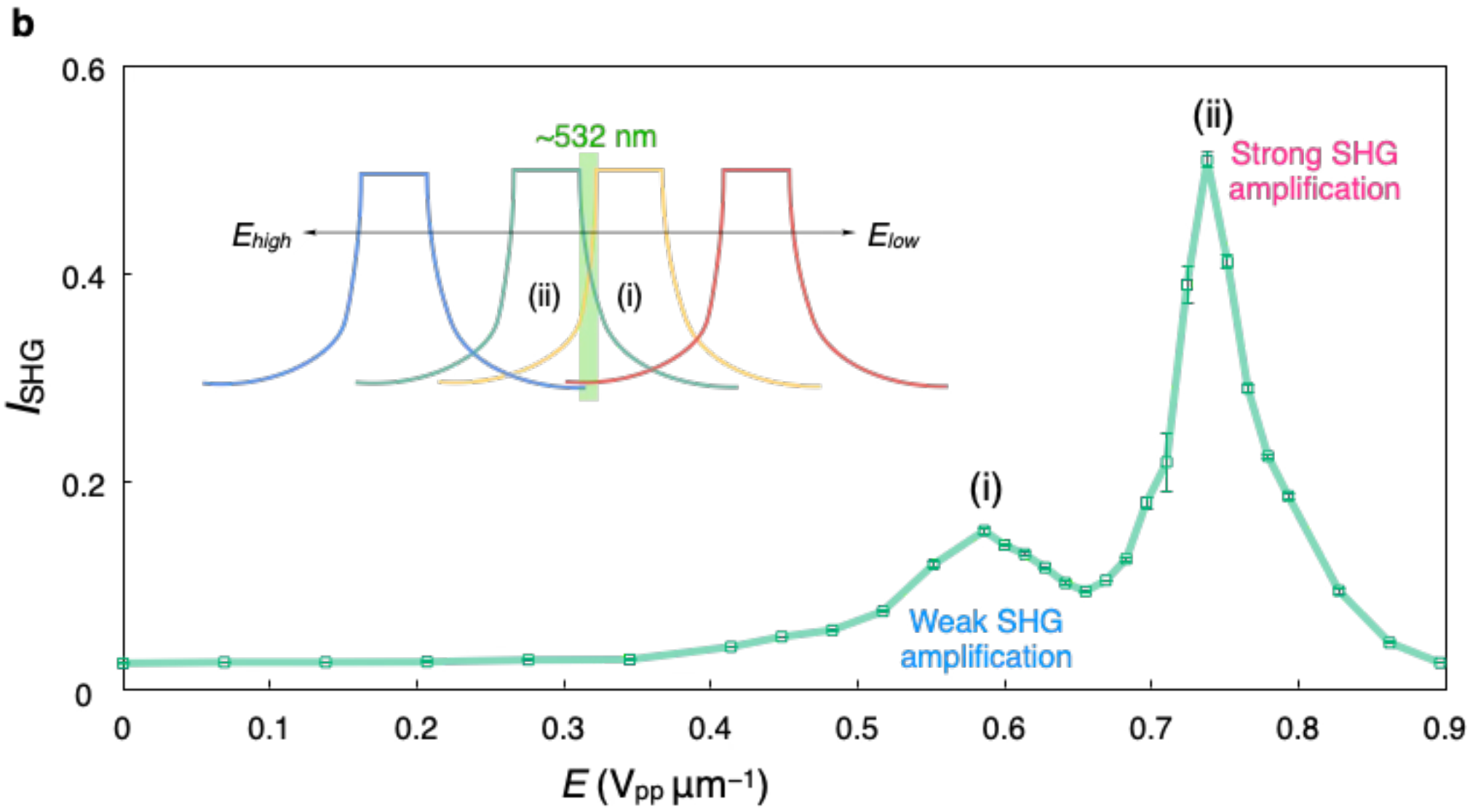

**Figure 6** SHG amplification in the $SmC_P^H$ phase: (a) schematic illustration and (b) SHG intensity vs *E*-field. The upper inset shows the ideal reflection band modulated by the *E*-field. The wavelength of the emitted SH light at ≈532 nm matches the band edge on the (i) short-wavelength or (ii) long-wavelength side.

# Supporting Information

## *Electrically Tunable Heliconical Smectic Superstructure in Polar Fluids*

Hiroya Nishikawa*, Dennis Kwaria, Atsuko Nihonyanagi, Koki Sano, Hiroyuki Yoshida and Fumito Araoka*

*To whom correspondence should be addressed.

E-mail: hiroya.nishikawa@riken.jp (H.N.), fumito.araoka@riken.jp (F.A.)

## Methods

### 1. General and materials

**Nuclear magnetic resonance (NMR) spectroscopy**: $^1$H, $^{13}$C, and $^{19}$F NMR spectra were recorded on JNM-ECZ500 (JEOL) operating at 500 MHz, 126 MHz, and 471 MHz for $^1$H [$^1$H{$^{19}$F}], $^{13}$C{$^1$H} [$^{13}$C{$^1$H,$^{19}$F}] and $^{19}$F [$^{19}$F{$^1$H}] NMR, respectively, using the TMS (trimethylsilane) as an internal standard for $^1$H NMR and the deuterated solvent for $^{13}$C NMR. The absolute values of the coupling constants are given in Hz, regardless of their signs. Signal multiplicities were abbreviated by s (singlet), d (doublet), t (triplet), q (quartet), quint (quintet), sext (sextet), and dd (double–doublet), respectively.

**High-resolution mass (HRMS) spectroscopy**: The high-resolution electrospray ionization mass spectrometry (HRESI-MS) was performed on SYNAPT G2-S/UPLC system (Waters ltd.).

**Density Functional Theory (DFT) Calculation**: Calculations were performed using the Chem3D (pro, 22.2.0.3300) and Gaussian 16 (G16, C.01) softwares (installed at the RIKEN Hokusai GreatWave Supercomputing facility) for MM2 and DFT calculations, respectively. GaussView 6 (6.0.16) software was used to visually analyze the calculation results. [S1] Dipole moments of molecules were calculated using b3lyp-gd3bj/6-311+g(d,p) level. The calculation method is as follows: opt=tight b3lyp-gd3bj/6-311+g(d,p) geom=connectivity empiricaldispersion=gd3bj int=ultrafine.

**Polarized optical microscopy (POM)**: Polarized optical microscopy was performed on a polarizing microscope (Eclipse LV100 POL, Nikon) with a hot stage (HSC402, INSTEC) on the rotation stage. Unless otherwise noted, the sample temperature was controlled using the INSTEC temperature controller (mk2000, INSTEC).

**Differential scanning calorimetry (DSC).** Differential scanning calorimetry was performed on a calorimeter (DSC3+, Mettler-Toledo). Rate: 10 K min$^{-1}$. Cooling/heating profiles were recorded and analyzed using the Mettler-Toledo STARe software system.

**Dielectric spectroscopy.** Dielectric relaxation spectroscopy was performed ranging between 1 Hz and 1 MHz using an impedance/gain-phase analyzer (SI 1260, Solartron Metrology) and a dielectric interface (SI 1296, Solartron Metrology). Prior to starting the measurement of the LC sample, the capacitance of the empty cell was determined as a reference. The temperature was controlled using a homemade heater block and temperature controller. The automatic measurement was taken place by the LabVIEW software programming.

**PRC measurement.** PRC measurements were performed in the temperature range of the polar phase under a triangular-wave *E*-field using a ferroelectricity evaluation system (FCE 10, TOYO Corporation), which is composed of an arbitrary waveform generator (2411B), an IV/QV amplifier (model 6252) and a simultaneous A/D USB device (DT9832).

**SHG/SHM measurements.** The SHG/SHM investigation was carried out using a Q-switched DPSS Nd: YAG laser (FQS-400-1-Y-1064, Elforlight) at $\lambda$ = 1064 nm with a 5 ns pulse width (pulse energy: 400 μJ). The primary beam was incident on the LC cell followed by the detection of the SHG signal. The electric field was applied normally to the LC cell.

**Wide-angle X-ray diffraction (WAXD) analysis.** 2D WAXD measurement was performed using NANOPIX system (Rigaku). The samples held in a homemade holder with magnets [S12] was measured at a constant temperature using a temperature controller and a hot stage (mk2000, INSTEC) with high temperature-resistance samarium cobalt magnets (~1 T, Shimonishi Seisakusho Co., Ltd.). The scattering vector $q$ ($q = 4\pi\sin\theta\ \lambda^{-1}$; $2\theta$ and $\lambda$ = scattering angle and wavelength of an incident X-ray beam (1.54 Å) and position of an incident X-ray beam on the detector were calibrated using several orders of layer diffractions from silver behenate ($d$ = 58.380 Å). The sample-to-detector distances were 82.20 mm, where acquired scattering 2D images were integrated along the Debye–Scherrer ring by using software (Igor Pro with Nika-plugin), affording the corresponding one-dimensional X-ray diffraction profiles.

**Information of used liquid crystalline (LC) cells**:

*ITO glass cell (EHC)*:

- ITO-coated type, electrode area: 5 × 10 mm
- Experiments: POM (thickness: 10.0 and 25.0 μm) and DR (thickness: 10.0 μm)

*ITO glass cell (homemade)*:

(case i) ITO-coated type (non-alignment layer), electrode area: 4 × 5 mm

(case i) Experiments: PRC (thickness: 4.6 μm) studies

(case ii) ITO-coated type (silanized layer), electrode area: 4 × 5 mm

(case ii) Experiments: Pitch modulation (thickness: 14.1 μm) studies

*Synparallel-rubbed (SP) cell (EHC)*:

- PI-coated type, thickness: 5.0 μm
- Alignment layer: LX-1400
- Experiments: POM studies

## 2. Synthesis of Compound 1.

### 2.1. Synthetic route

All compounds (**1–6**) were synthesized according to previous literature [S2] (Scheme S1). NMR and MS spectra for final target **1** are shown in Figure S28–S31.

**Scheme S1** Synthetic pathway of **Compound 1**. a) $B_2pin_2$, Pd(dppf)$Cl_2$-$CH_2Cl_2$, KOAc, 1,4-dioxane, MM400 (30 Hz), 110 °C, 10 min, b) Pd$(OAc)_2$, SPhos, $K_2CO_3$, THF/$H_2O$, 55 °C, 3 h, c) 3,4-dihydro-2*H*-pyran, $Et_2O$, 35 °C, 2.5 h; r.t., 20 h, d) **3**, EDAC-HCl, DMAP, $CH_2Cl_2$, 0 °C, 1 h; r.t., 1 h, e) *p*-TsOH, $CH_2Cl_2$/MeOH, 50 °C, 1.5 h, f) TEA, $CH_2Cl_2$, 0 °C; r.t.

**2.2.1.** ***Synthesis of 2-fluoro-4-((tetrahydro-2H-pyran-2-yl)oxy)benzoic acid*** **(4)**

$^{1}$H{$^{19}$F}-NMR (500 MHz, $CDCl_3$): δ 7.97 (d, *J* = 8.7 Hz, 1H), 6.90–6.86 (m, 2H), 5.50 (t, *J* = 2.7 Hz, 1H), 3.85–3.80 (m, 1H), 3.65 (dt, *J* = 11.3, 3.9 Hz, 1H), 2.04–1.95 (m, 1H), 1.90–1.87 (m, 2H), 1.77–1.67 (m, 2H), 1.63–1.60 (m, 1H)

$^{19}$F{$^{1}$H}-NMR (471 MHz, $CDCl_3$): δ −105.0

**2.2.2.** ***Synthesis of 4'-(difluoro(3,4,5-trifluorophenoxy)methyl)-2,3',5'-trifluoro-[1,1'-biphenyl]-4-yl 2-fluoro-4-((tetrahydro-2H-pyran-2-yl)oxy)benzoate*** **(5)**

$^{1}$H{$^{19}$F}-NMR (500 MHz, $CDCl_3$): δ 8.04 (d, J = 8.7 Hz, 1H), 7.48 (d, *J* = 8.0 Hz, 1H), 7.22 (s, 2H), 7.18–7.16 (m, 2H), 7.00 (d, *J* = 5.6 Hz, 2H), 6.96-6.91 (m, 2H), 5.53 (t, *J* = 3.0 Hz, 1H), 3.86–3.81 (m, 1H), 3.68–3.64 (m, 1H), 2.04–1.96 (m, 1H), 1.92–1.90 (m, 2H), 1.76–1.68 (m, 2H), 1.65–1.61 (m, 1H)

$^{19}$F{$^{1}$H}-NMR (471 MHz, $CDCl_3$): δ −61.7 (t, *J* = 27.3 Hz, 2F), −104.1 (s, 1F), −110.4 (t, *J* = 27.6 Hz, 2F), −113.9 (s, 1F), −132.3 (d, *J* = 18.4 Hz, 2F), −163.0 (t, *J* = 21.9 Hz, 1F)

**2.2.3.** ***Synthesis of 4'-(difluoro(3,4,5-trifluorophenoxy)methyl)-2,3',5'-trifluoro-[1,1'-biphenyl]-4-yl 2-fluoro-4-hydroxybenzoate*** **(6)**

$^{1}$H{$^{19}$F}-NMR (500 MHz, $CDCl_3$): δ 8.03 (d, *J* = 8.7 Hz, 1H), 7.48 (d, *J* = 8.6 Hz, 1H), 7.22 (s, 2H), 7.18z–7.16 (m, 2H), 7.00 (d, *J* = 5.7 Hz, 2H), 6.75 (dd, *J* = 8.6, 2.4 Hz, 1H), 6.70 (d, *J* = 2.4 Hz, 1H), 5.76 (s, 1H)

$^{19}$F{$^{1}$H}-NMR (471 MHz, $CDCl_3$): δ -61.7 (t, J = 25.7 Hz, 2F), -104.0 (s, 1F), -110.3 (t, J = 25.9 Hz, 2F), −113.9 (s, 1F), −132.3 (d, *J* = 22.1 Hz, 2F), −163.0 (t, *J* = 22.1 Hz, 1F)

**2.2.4.** ***Synthesis of 4'-(difluoro(3,4,5-trifluorophenoxy)methyl)-2,3',5'-trifluoro-[1,1'-biphenyl]-4-yl 4-(butyryloxy)-2-fluorobenzoate (Compound 1)***

White powder, 84.9% yield

$^{1}$H{$^{19}$F}-NMR (500 MHz, $CDCl_3$): δ 8.14 (dd, *J* = 7.8, 1.5 Hz, 1H), 7.50 (dd, *J* = 7.9, 0.9 Hz, 1H), 7.23 (s, 2H), 7.20–7.18 (m, 2H), 7.10–7.08 (m, 2H), 7.00 (d, *J* = 5.6 Hz, 2H), 2.59 (t, *J* = 7.4 Hz, 2H), 1.81 (td, *J* = 14.8, 7.4 Hz, 2H), 1.07 (t, *J* = 7.4 Hz, 3H)

$^{19}$F{$^{1}$H}-NMR (471 MHz, $CDCl_3$): δ -61.7 (t, *J* = 25.7 Hz, 2F), −103.9 (s, 1F), −110.3 (t, *J* = 25.7 Hz, 2F), −113.6 (s, 1F), −132.3 (d, *J* = 18.4 Hz, 2F) , −163.0 (t, *J* = 21.9 Hz, 1F)

$^{13}$C{$^{1}$H,$^{19}$F}-NMR (126 MHz, $CDCl_3$): δ 171.0, 162.8, 161.5, 159.9, 159.5, 156.1, 151.9, 151.0, 144.6, 140.7, 138.5, 133.4, 130.6, 123.4, 120.1 (t, *J* = 262.5 Hz), 118.4, 117.8, 114.6, 113.1, 111.2, 110.9, 108.9 (t, *J* = 32.1 Hz), 107.5, 36.1, 18.2, 13.6

ESI-HRMS (m/z): [M+Na$^{+}$] calc for 651.0830; found, 651.0834

**Note 1 Characteristic of HEC phase**

The distinct features of the present “HEC” phase are summarized below:

(1) it exhibits a helielectric conical ordering and appears below the $N_{TBF}$ phase via a continuous transition;

(2) XRD measurements upon magnetic field application indicate a tilted smectic-like layering, but without long-range positional order, in contrast to the $N_{TBF}$ and $SmC_P^H$ phases; and

(3) the $SmC_P^H$ phase emerges from this phase through a distinct first-order phase transition.

Based on these observations, we initially presumed that the HEC phase corresponds to a helielectric conical structure with cybotactic SmC-like clusters, rather than a true smectic phase possessing long-range positional order. However, these findings alone are insufficient to conclusively classify it as a cluster phase. Therefore, we designate it as a helielectric conical phase (HEC phase). This terminology was also adopted in our previous work [S2] with the same rationale, and hence we use the same notation “HEC” here.

**Note 2 DC-bias-dependent low-frequency dielectric response evaluated by BDS.**

To examine whether mobile-charge-related effects contribute to the different responses under AC/DC driving, we performed DC-bias-dependent broadband dielectric spectroscopy (BDS) measurements using a 14-μm ITO cell. A small AC probing voltage of 100 mV was applied together with DC biases of 0–6 V, allowing the dielectric response of the DC-biased states to be evaluated. The complex dielectric spectra were analyzed using two Havriliak–Negami relaxation contributions together with the conductivity term employed in a previous dielectric study of polar LCs [S13]:

$$\varepsilon^*(f) = \varepsilon_\infty + \sum_{k=1}^{2} \frac{\Delta\varepsilon_k}{\left[1 + \left(j\frac{f}{f_k}\right)^{a_k}\right]^{b_k}} + \left(\frac{\sigma_{DC}}{j2\pi f \varepsilon_0}\right)^N$$

, where $\varepsilon_\infty$, $\Delta\varepsilon_k$ and $f_k$ are the high-frequency permittivity, the dielectric strength and relaxation frequency, respectively; $a_k$ and $b_k$ are broadness exponents of the corresponding relaxation mode; $\sigma_{DC}$ denotes the DC conductivity parameter in the low-frequency term, while $N$ is an exponent with values between 0 and 1. For a consistent comparison among different DC biases, $b_k$ was fixed at 1 and $N$ was treated as a common global parameter for all spectra, yielding $N = 0.897$. Because this fitted value is slightly below unity, the field dependence of this contribution is discussed in terms of a low-frequency charge-related response, rather than as a direct measure of intrinsic bulk DC conductivity. The fitted BDS spectra and the DC-field dependence of the effective conductivity evaluated at 1 Hz from the fitted conductivity term are shown in Figures S23 and S24, respectively. The extracted low-frequency charge-related response under DC bias increased markedly, reached an almost saturated level by 0.286 V μm$^{-1}$, and subsequently decreased at higher static fields. Importantly, the DC electric fields of 0.286 and 0.357 Vμm$^{-1}$ directly overlap with the AC field range in which pronounced optical pitch modulation was observed, 0.23–0.39 $V_{rms}$ μm$^{-1}$. Therefore, the low-frequency charge-related response under DC bias is already nearly saturated in the field range where AC driving produces pronounced pitch modulation. Although these results do not directly visualize charge accumulation at the interfaces, they provide frequency-dependent dielectric evidence consistent with the development of interfacial charge redistribution and internal-field screening in the DC-biased state. Such a persistent static screening state can reduce the effective internal DC field acting on the $\mathrm{SmC_P^H}$ heliconical structure and thereby limit further pitch modulation under DC driving.

## Supporting Figures (Figures S1–S30)

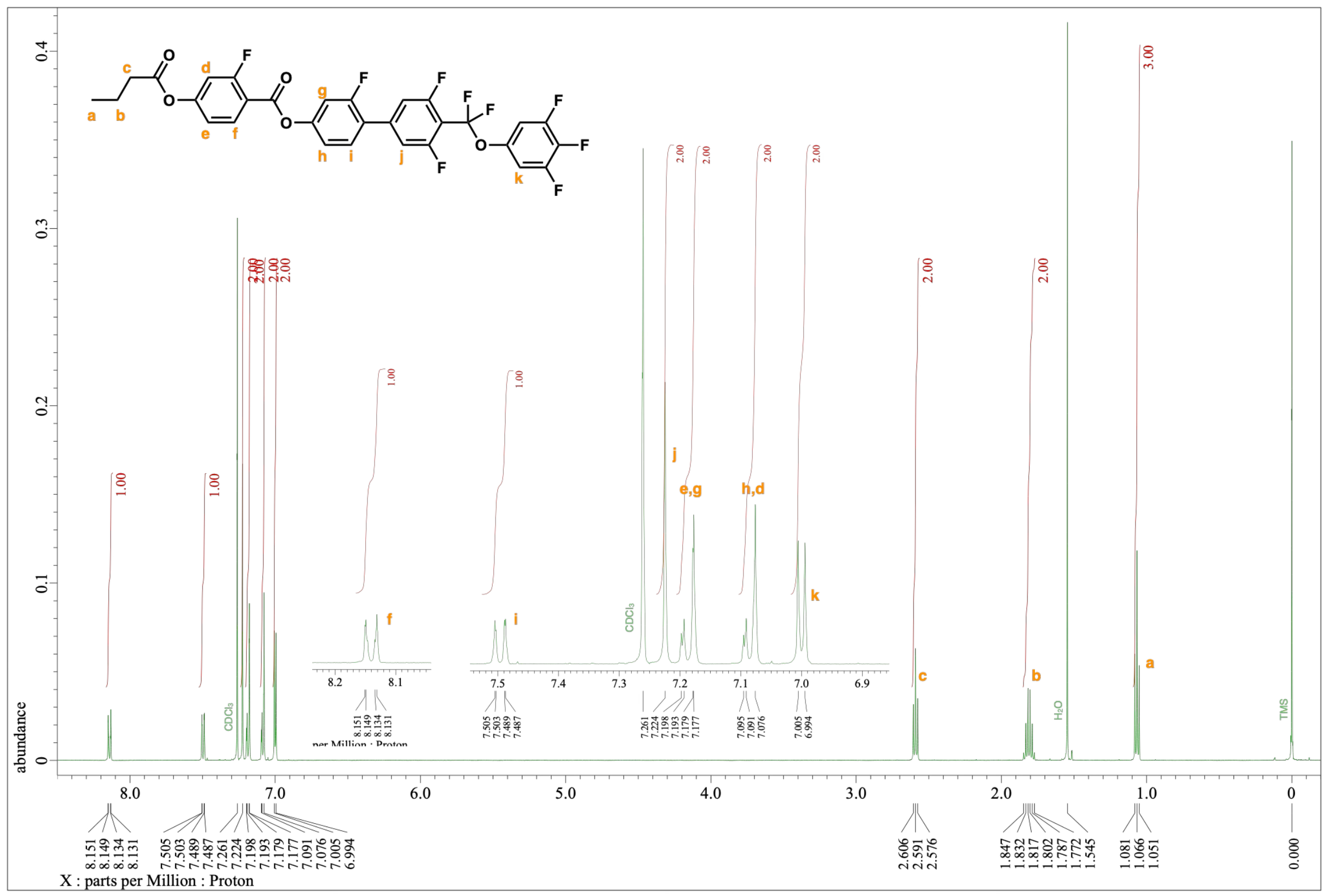

**Figure S1** $^{1}H\{^{19}F\}$NMR spectra for **Compound 1**.

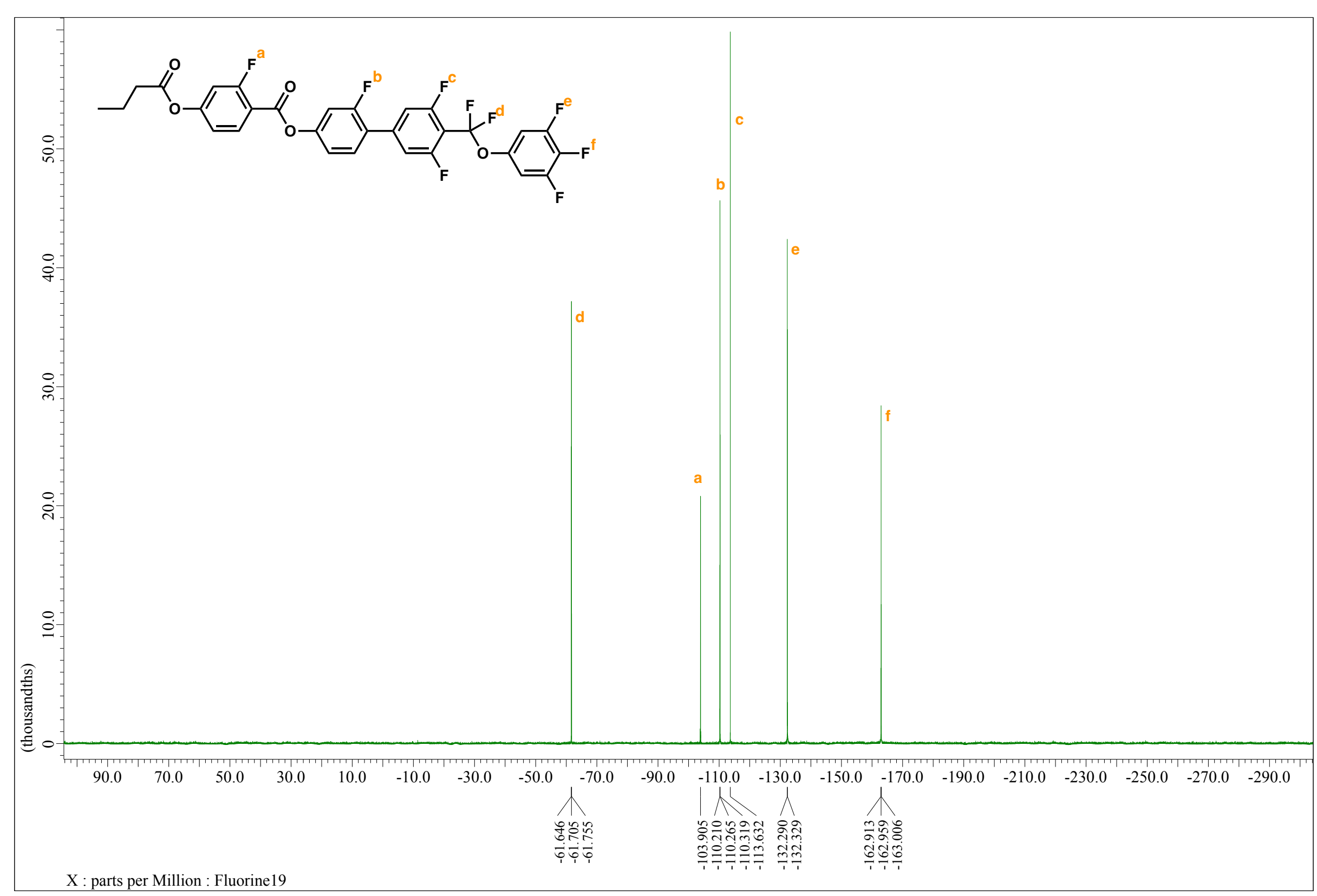

**Figure S2** $^{19}F\{^{1}H\}$NMR spectra for **Compound 1**.

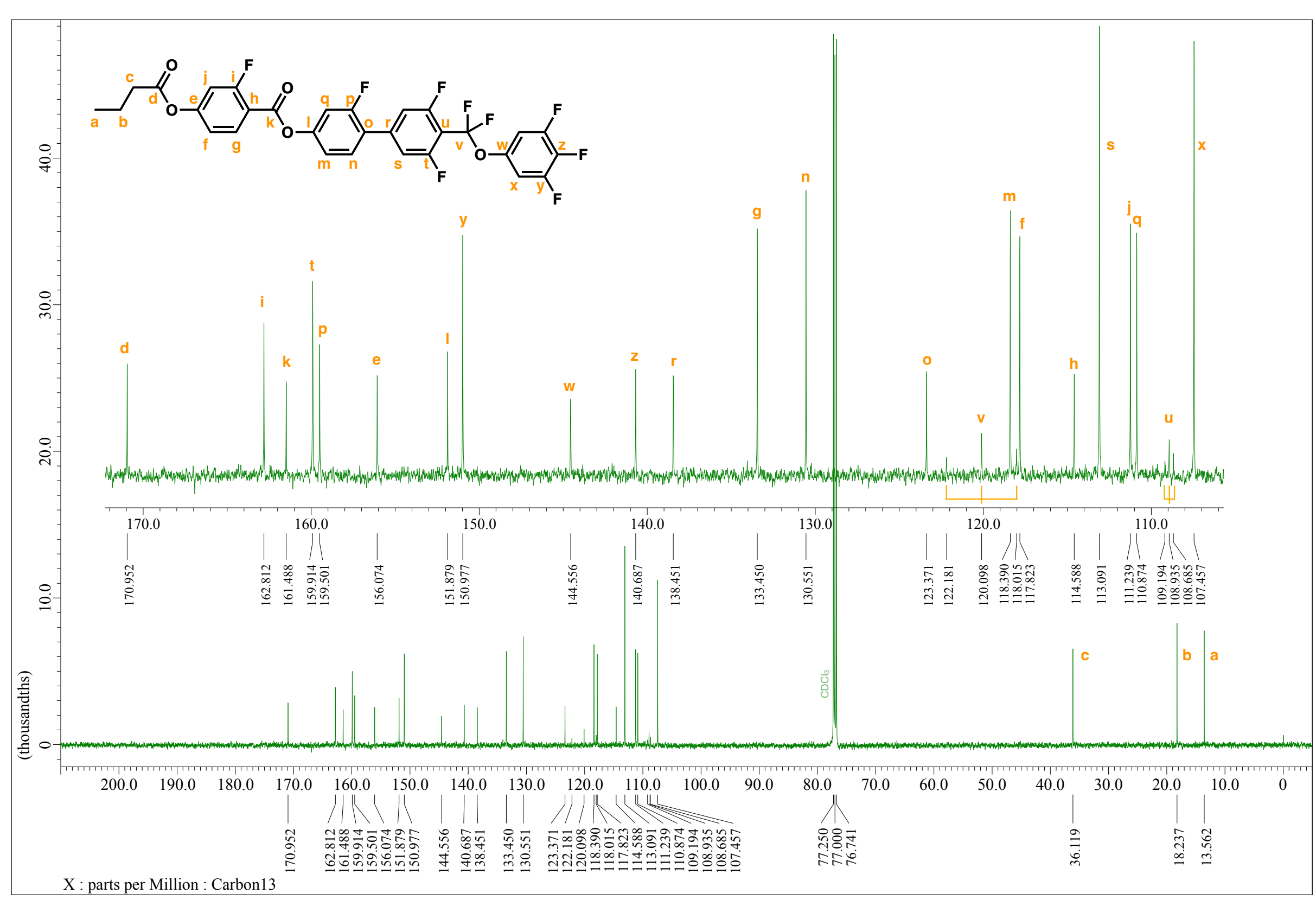

**Figure S3** $^{13}C\{^{1}H,^{19}F\}$NMR spectra for **Compound 1**.

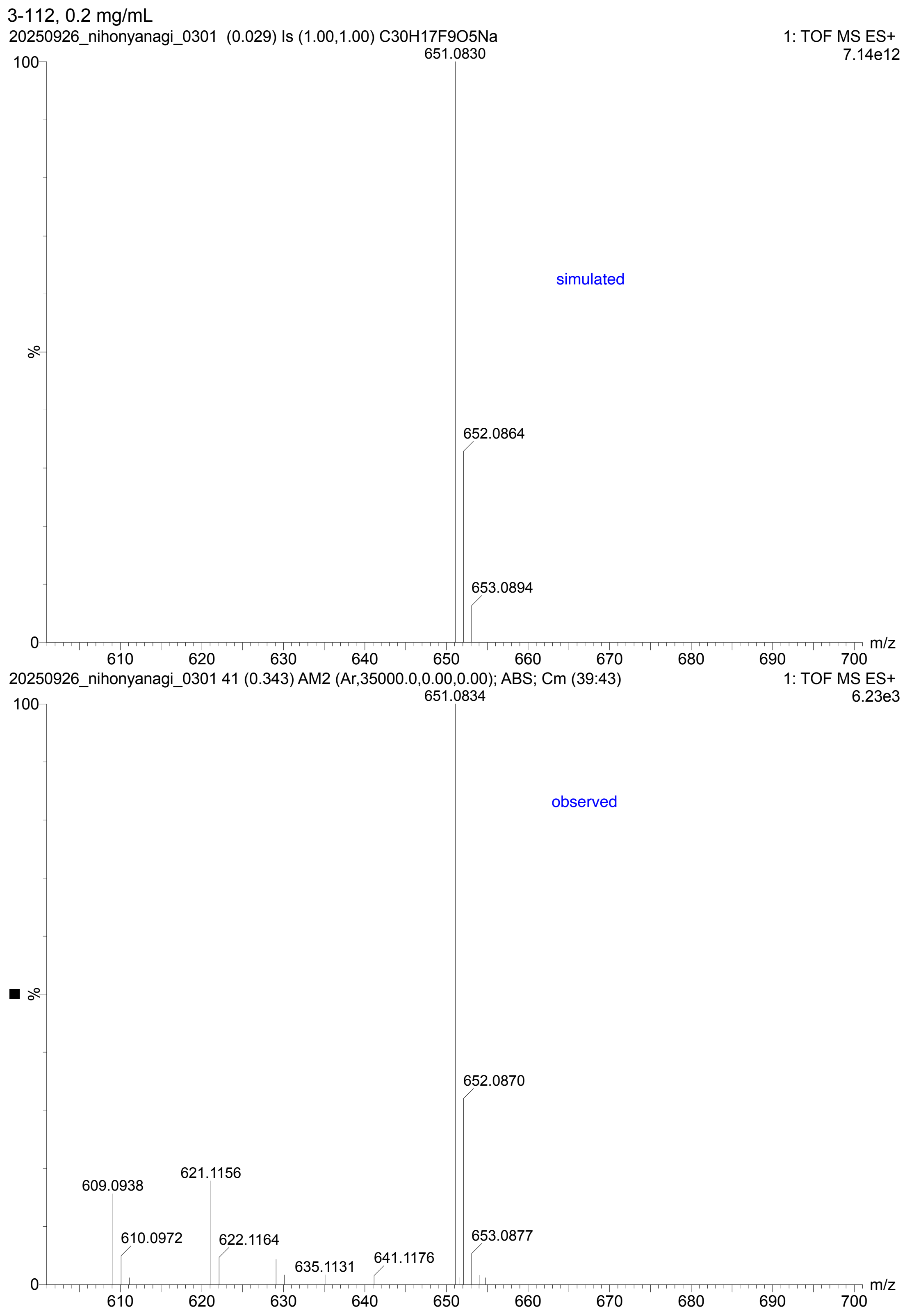

**Figure S4** HRMS spectra for **Compound 1**.

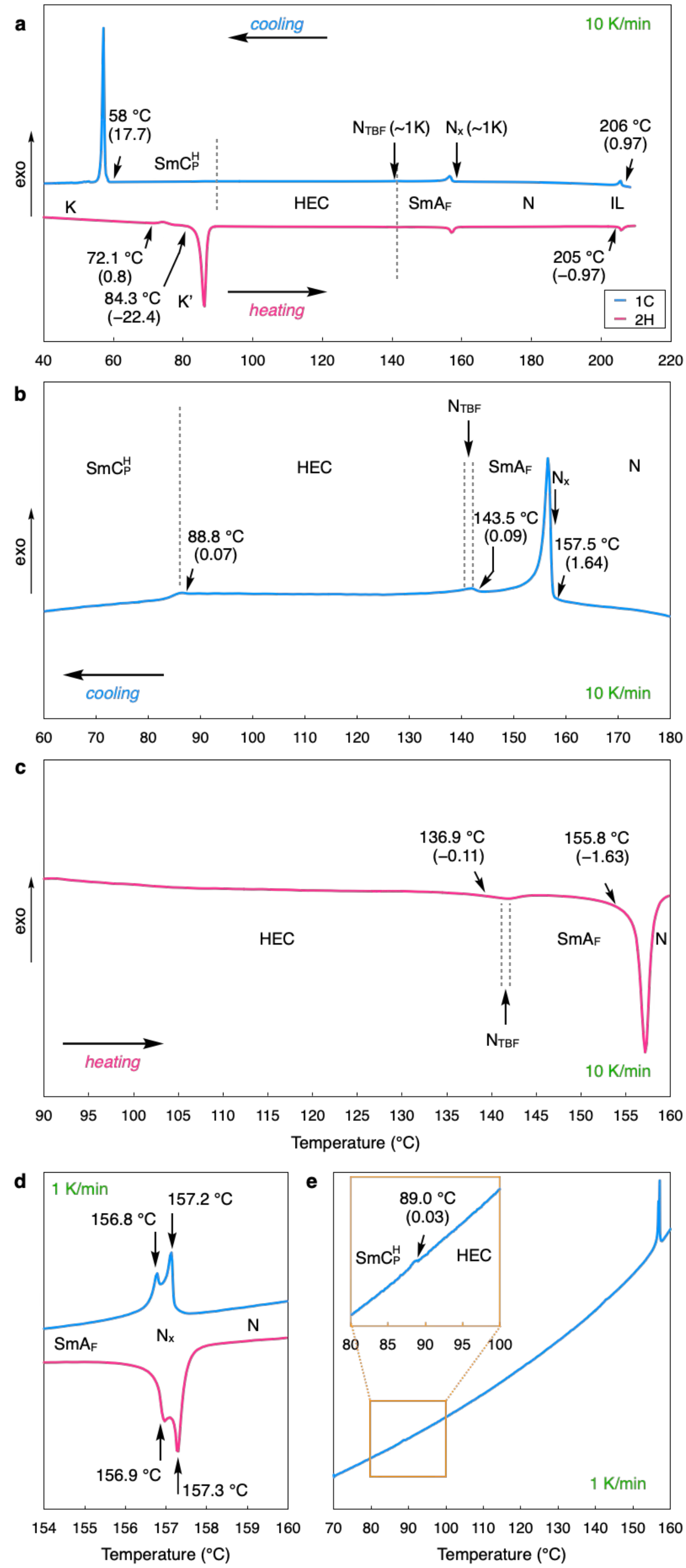

**Figure S5** DSC curves for **Compound 1**. Overall curves (a) and enlarged curves during cooling (b) and heating (c). Scan rate: 10 K min$^{-1}$. d,e) DSC curves recorded at a scan rate of 1 K min$^{-1}$.

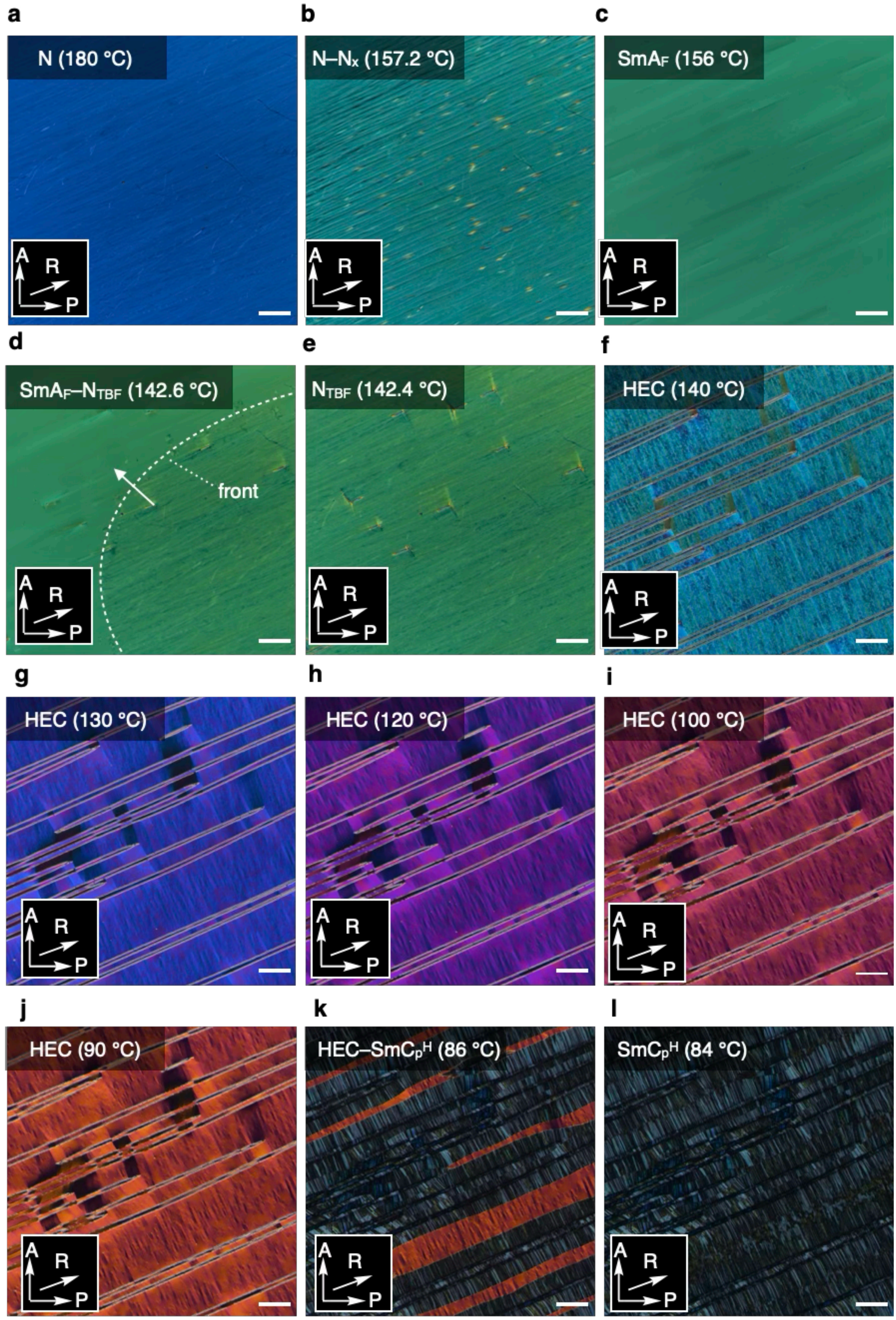

**Figure S6** Complete POM images taken in various temperature in SP cell (thickness: 5 μm) for **Compound 1**. Scale bar: 100 μm.

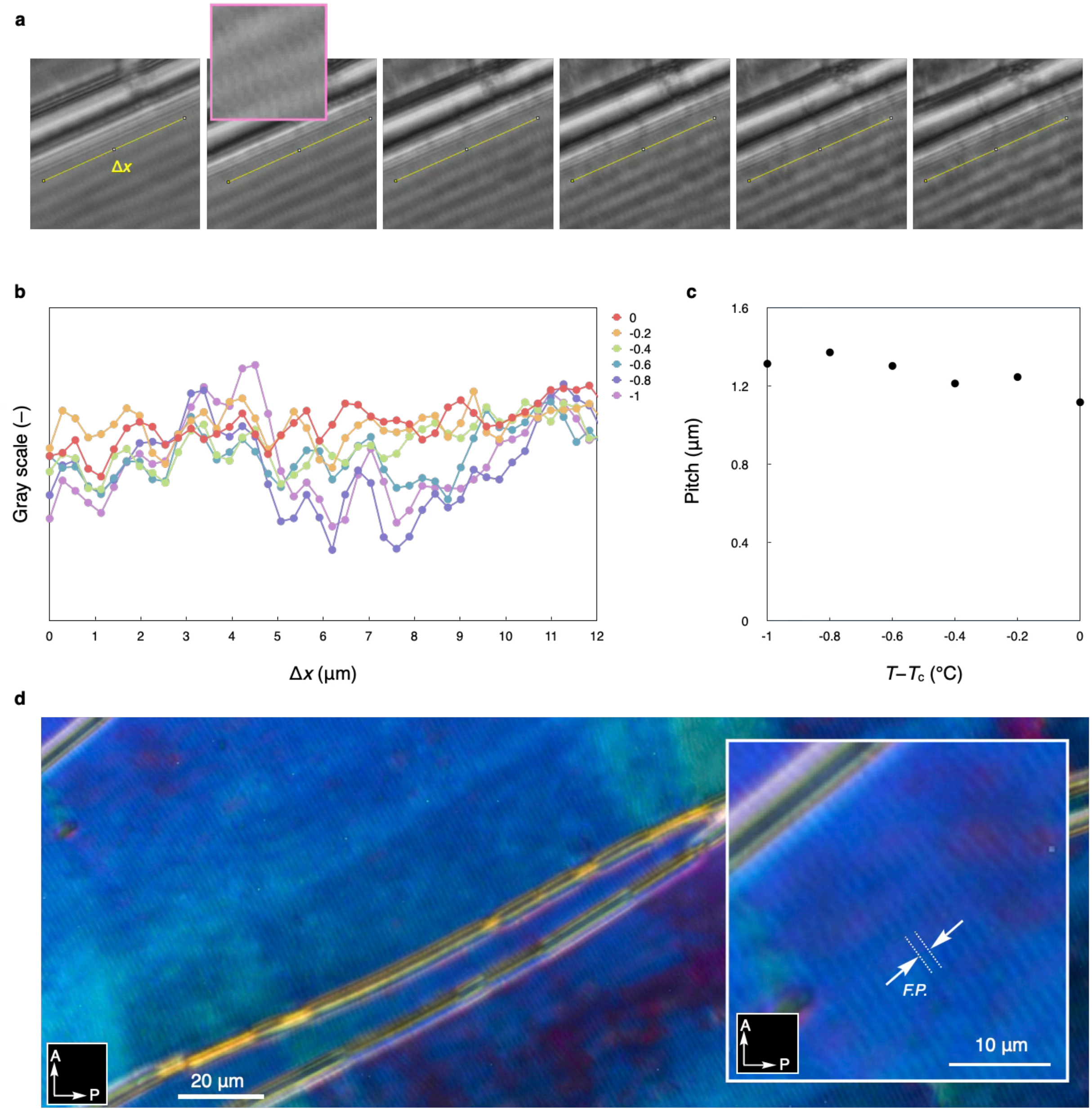

**Figure S7** Image analysis to estimate helical pitch in the $N_{TBF}$ range. a) 8-bit POM images, b) cross-sectional intensity profile perpendicular to fine stripes, c) estimated pitch vs temperature. d) POM image taken in the cover glass substrate cell without alignment layer and spacers. Similarly, the full pitch 1.3 μm was estimated.

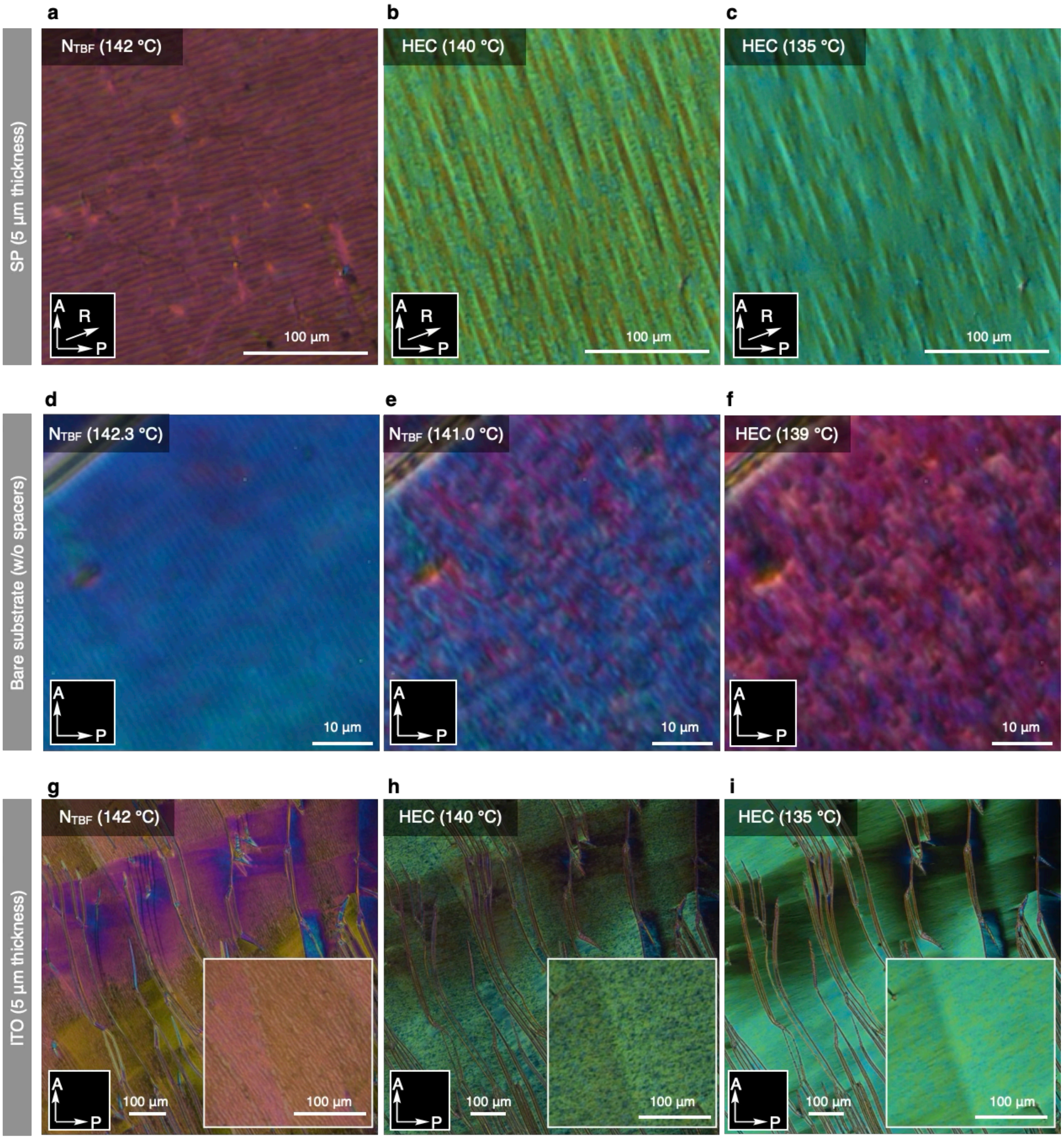

**Figure S8** Changes in POM texture taken in various temperature in SP (another lot) (a–c), bare substrate without spacers (d–f) and ITO cells (g–h) for **Compound 1**. Thickness: SP cell, 5.0 μm; ITO cell, 5.0 μm.

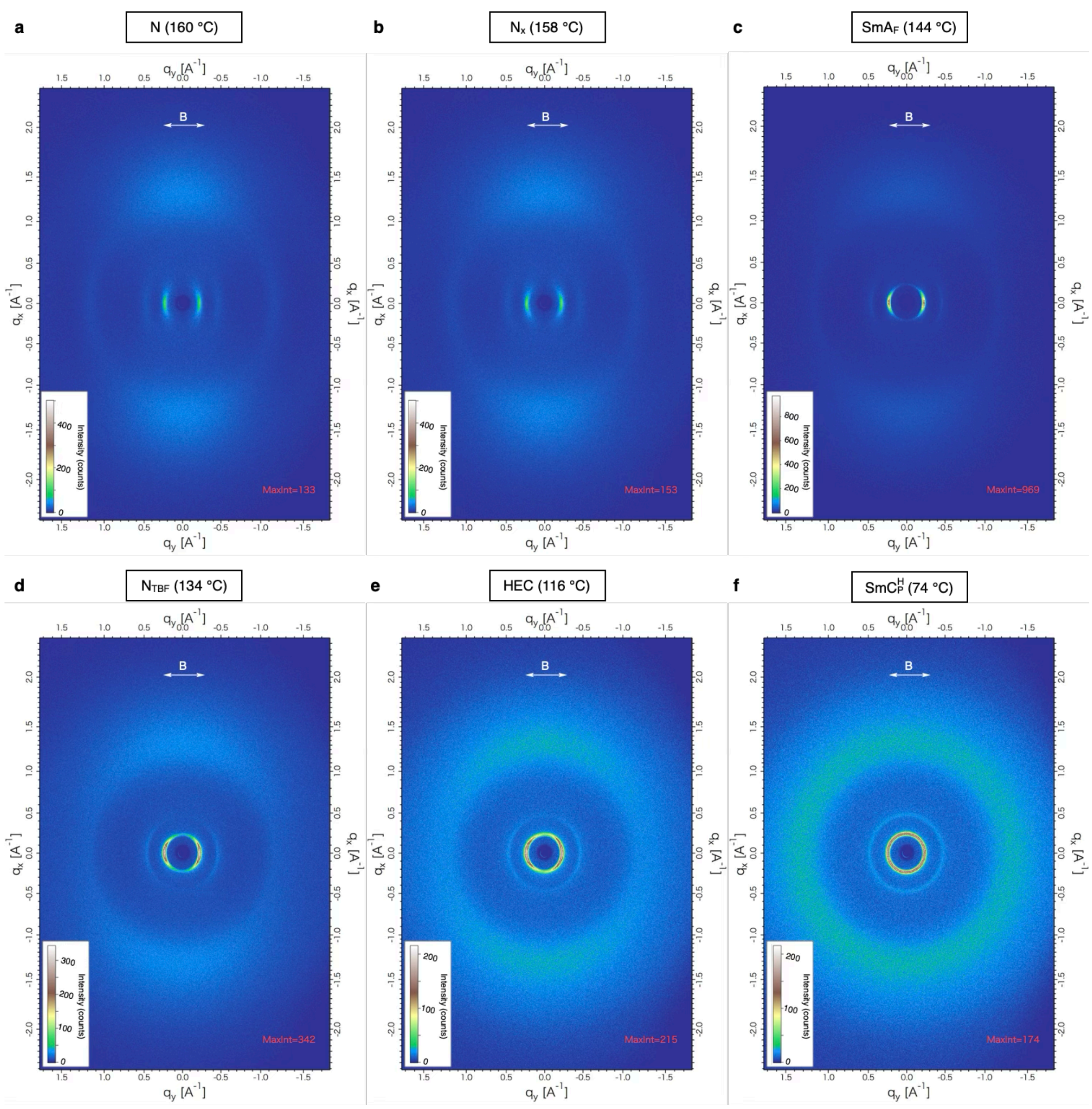

**Figure S9** Full-range 2D WAXD patterns in various phases under the magnetic field (~1 T, **B ∥ n**) for **Compound 1**.

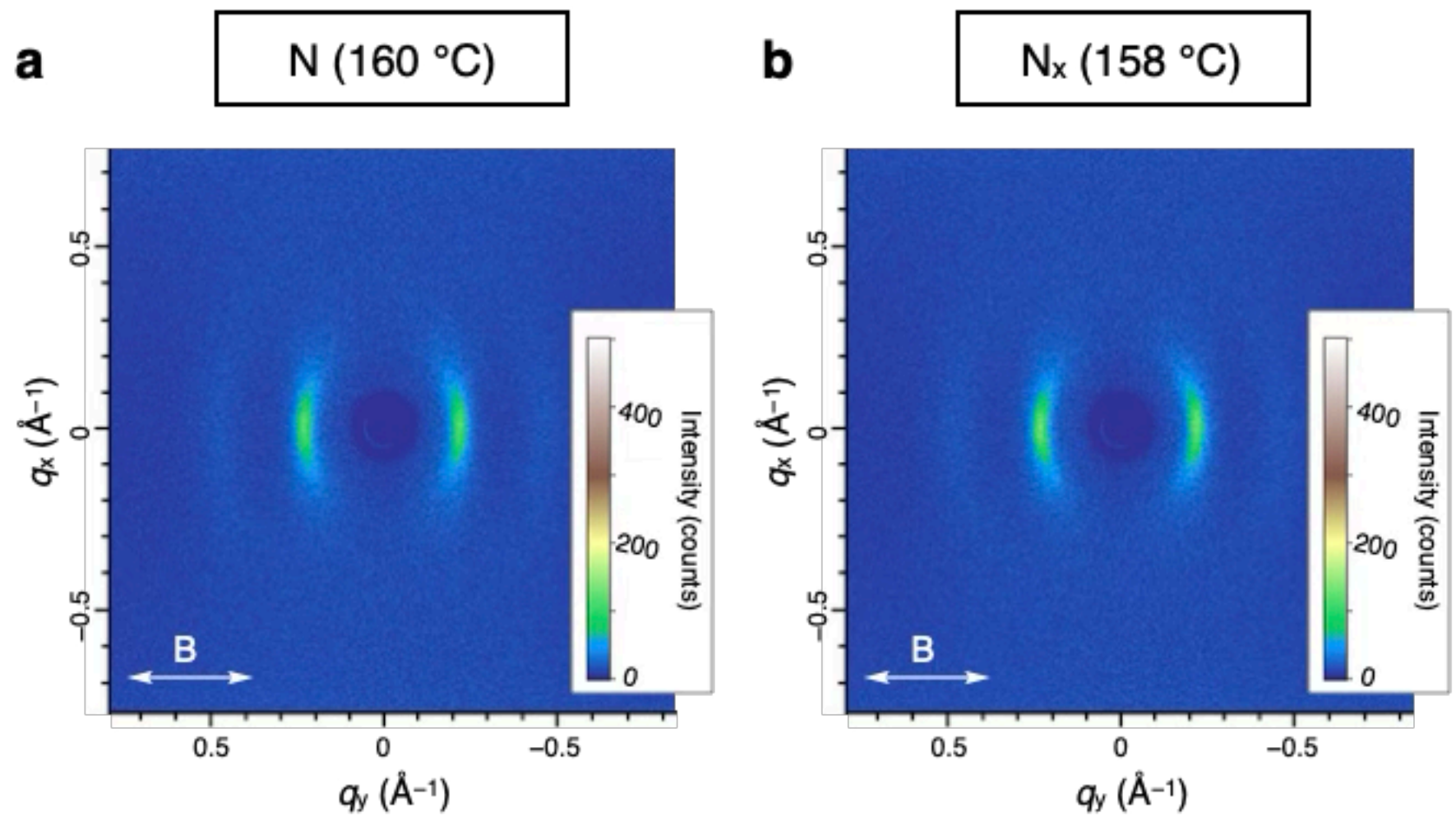

**Figure S10** 2D WAXD patterns in the N (a) and $N_x$ (b) phase under the magnetic field (~1 T, **B ∥ n**) for **Compound 1**.

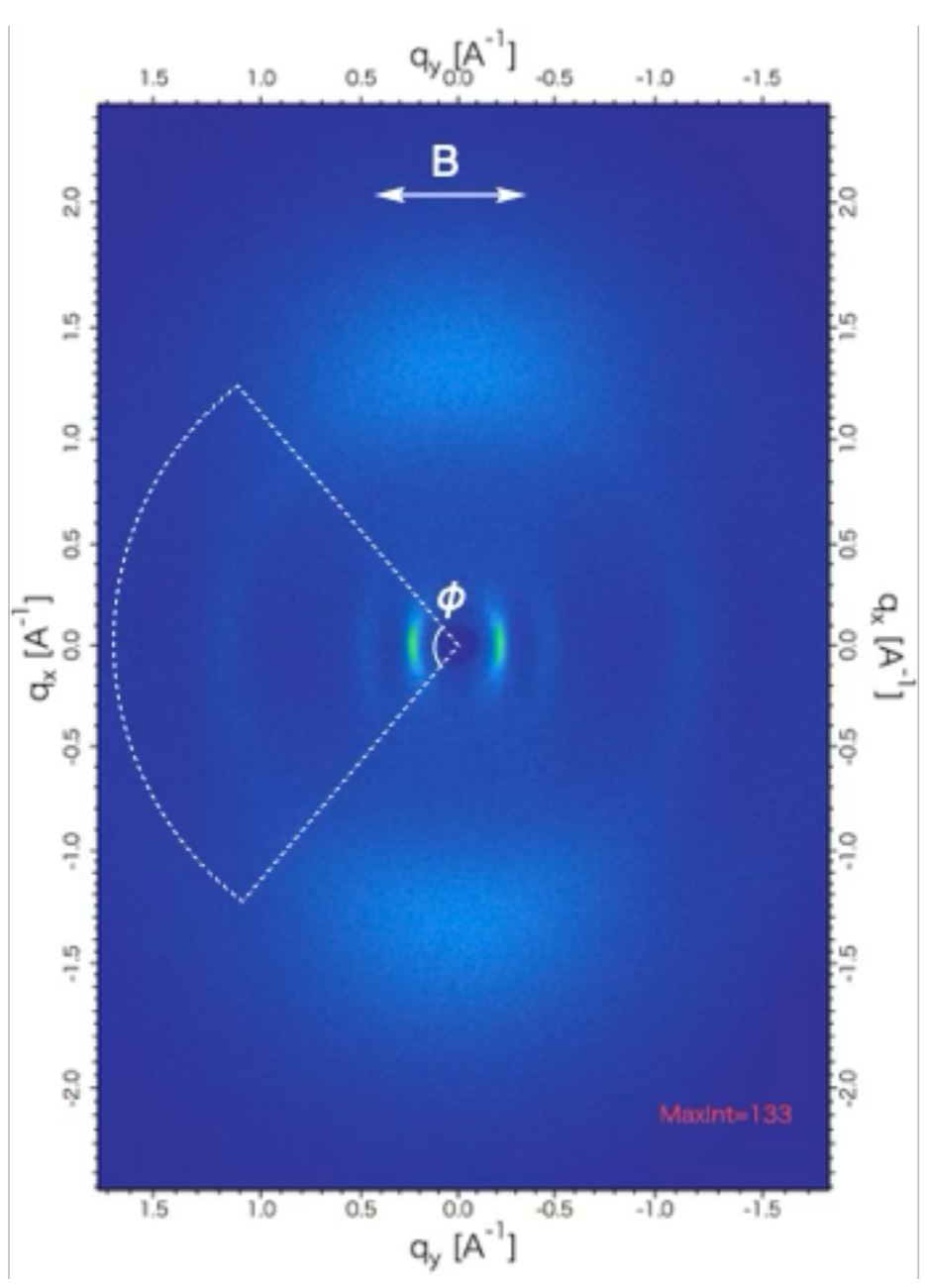

**Figure S11** Box scan area ($\varphi$ = 100°) on 2D X-ray pattern for generating 1D X-ray diffractogram.

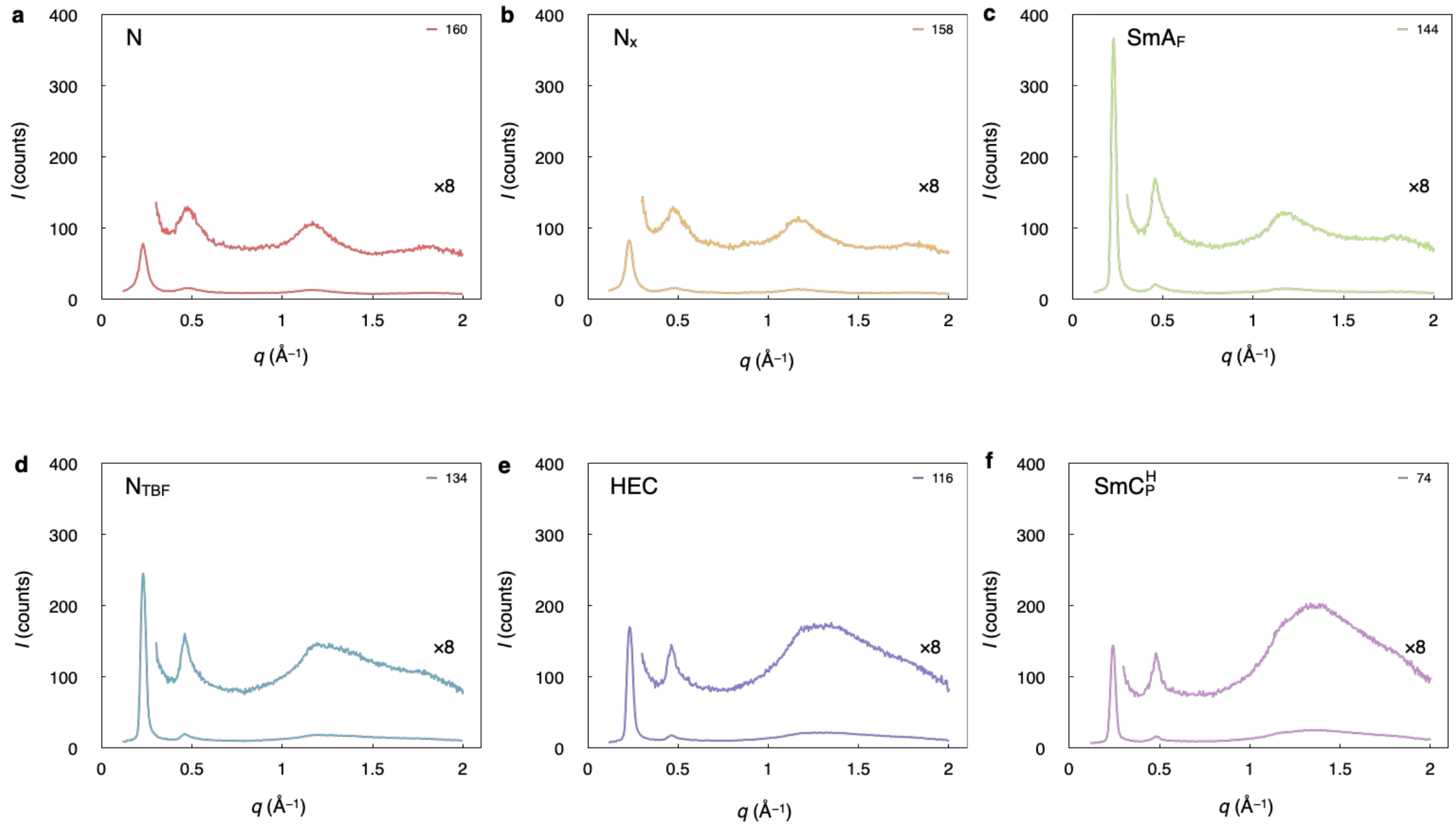

**Figure S12** 1D X-ray diffractogram along the equator direction (**n** ∥ **B**) from a box scan area ($\varphi$ = 100°).

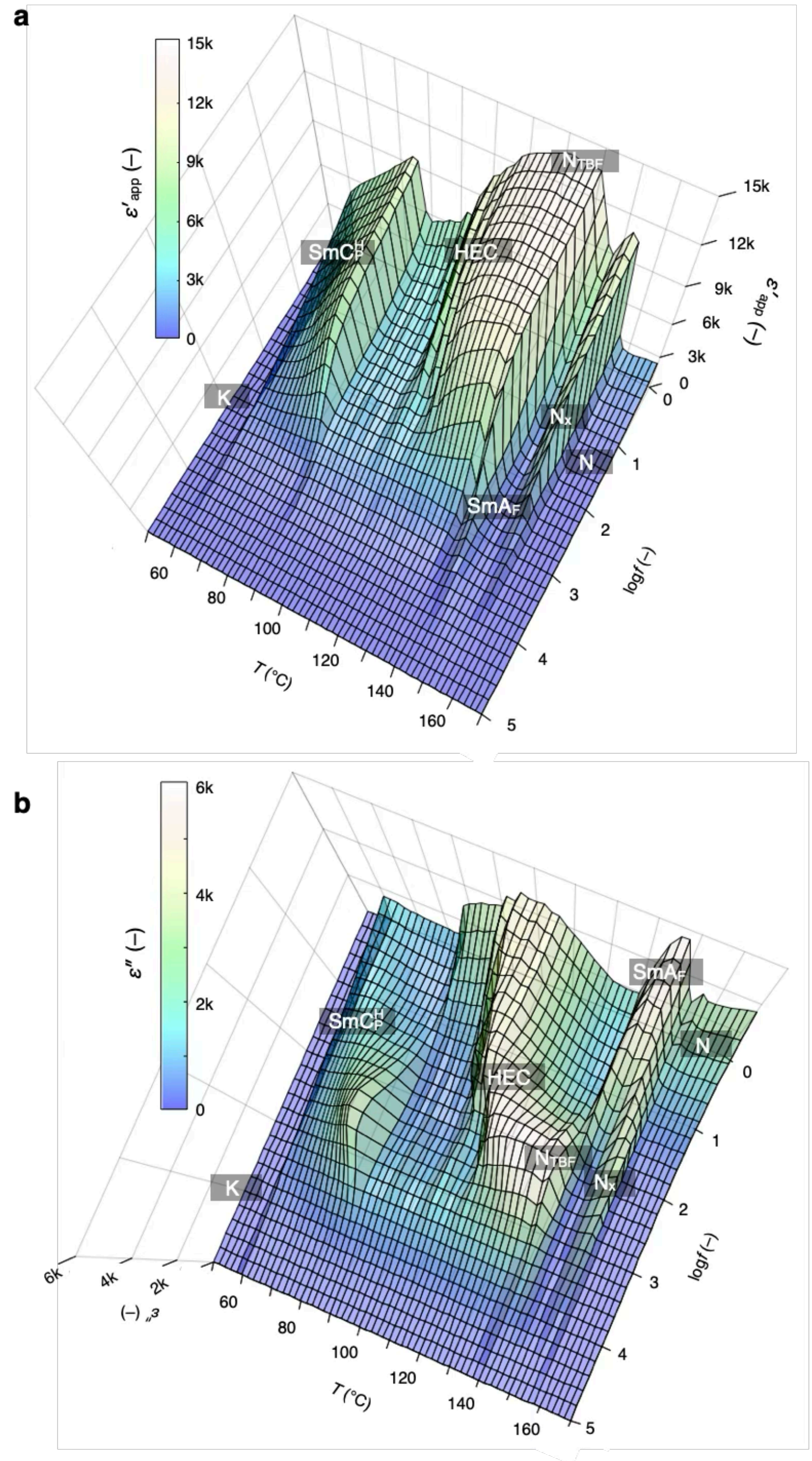

**Figure S13** 3D BDS data for **Compound 1** in various temperatures. (a) Apparent dielectric permittivity vs log*f*, Temperature; (b) dielectric loss vs log*f*, Temperature.

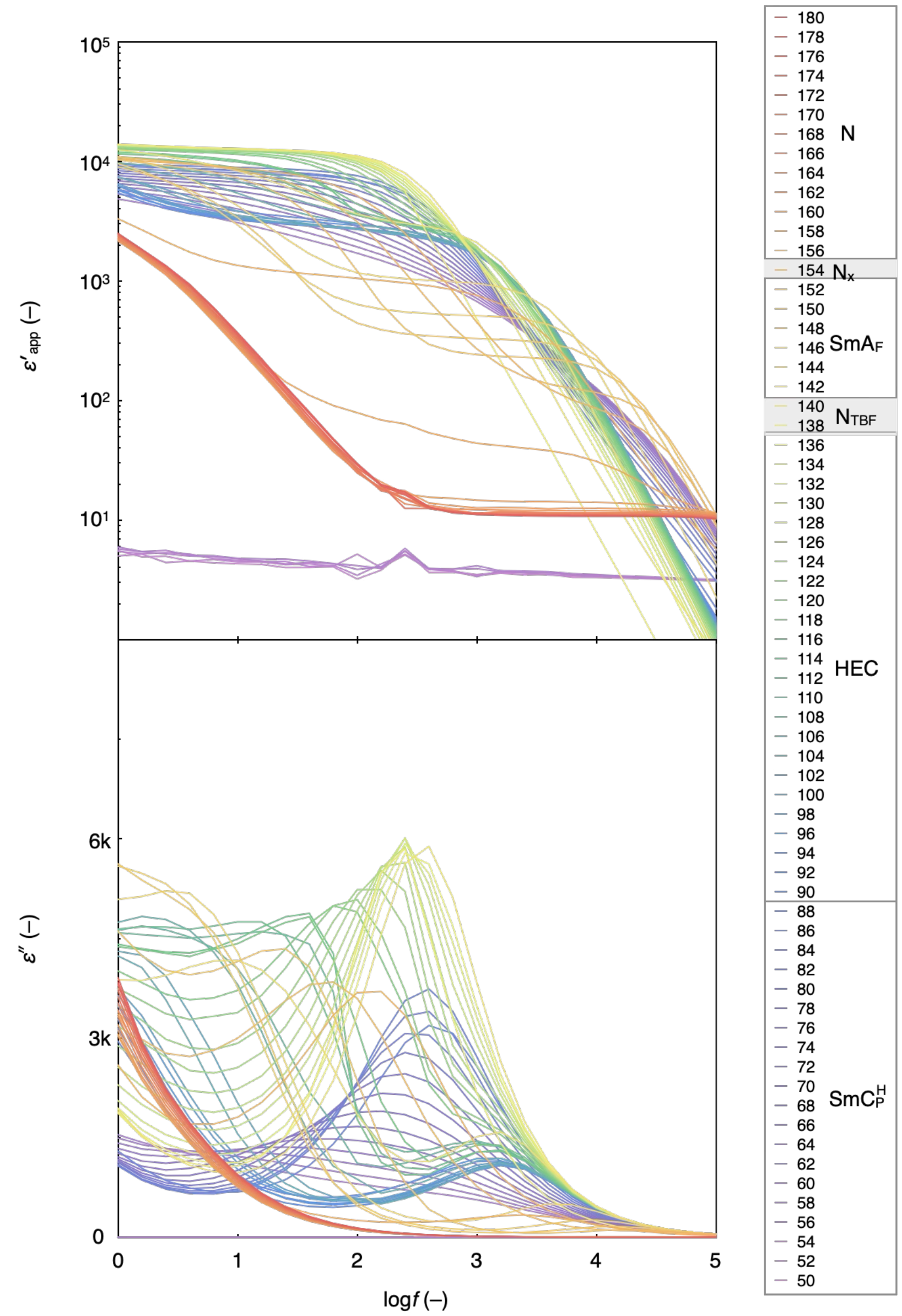

**Figure S14** Complete BDS data for **Compound 1** in various temperatures.

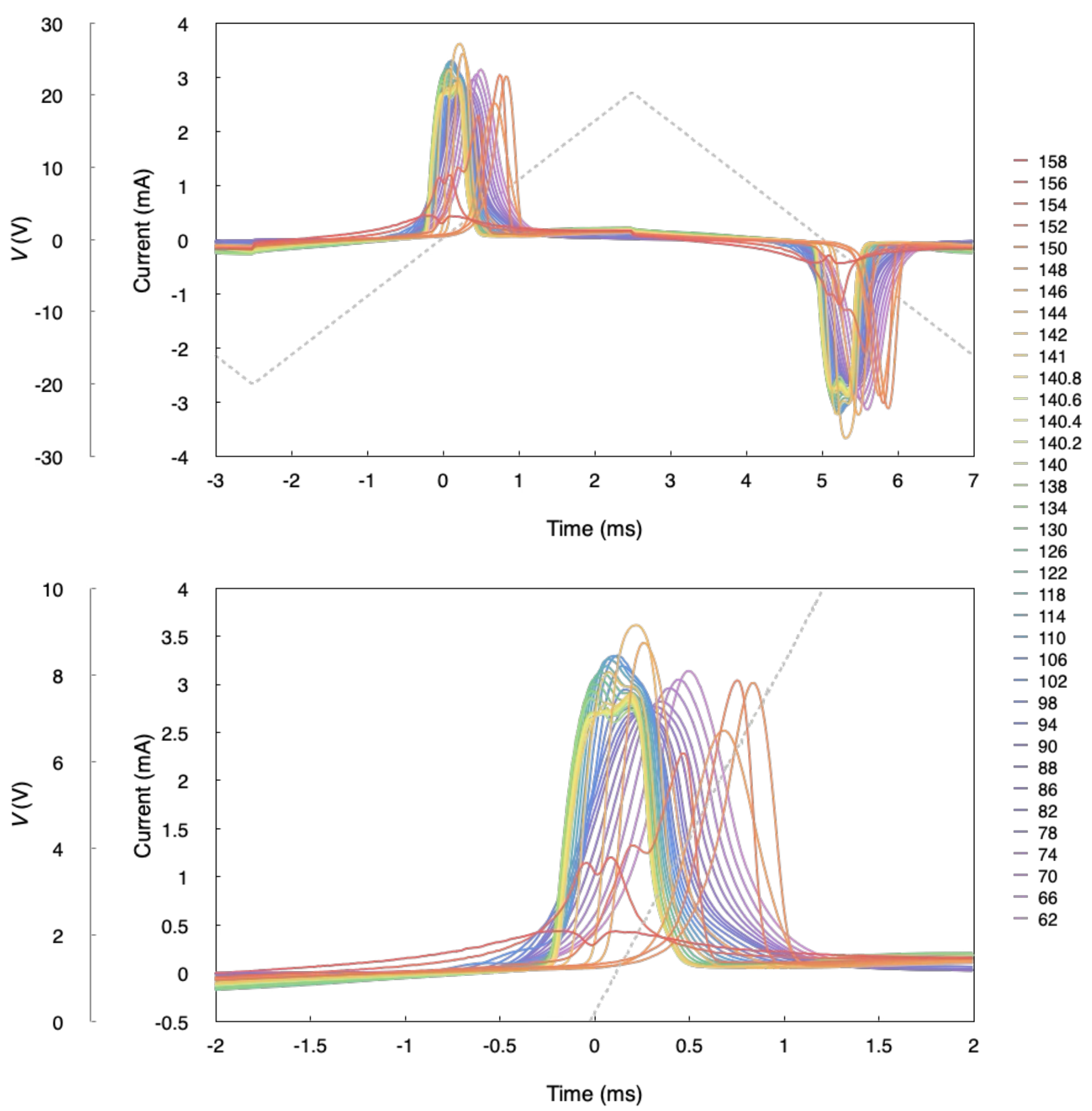

**Figure S15** Complete PRC data in various temperatures for **Compound 1**. Overall profile (a) and enlarged one (b).

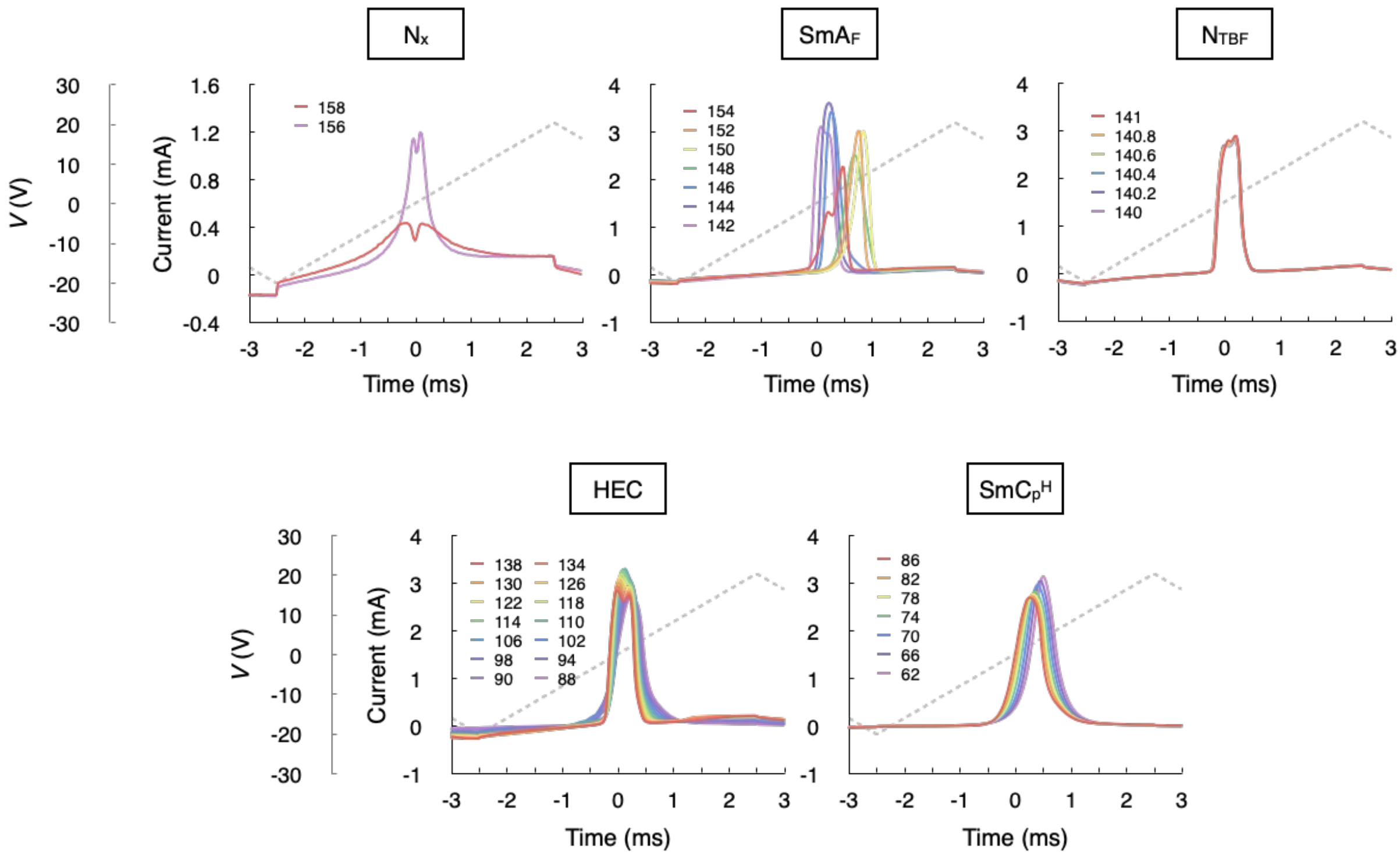

**Figure S16** Complete PRC data with applied voltage profile in various LC regions for **Compound 1**.

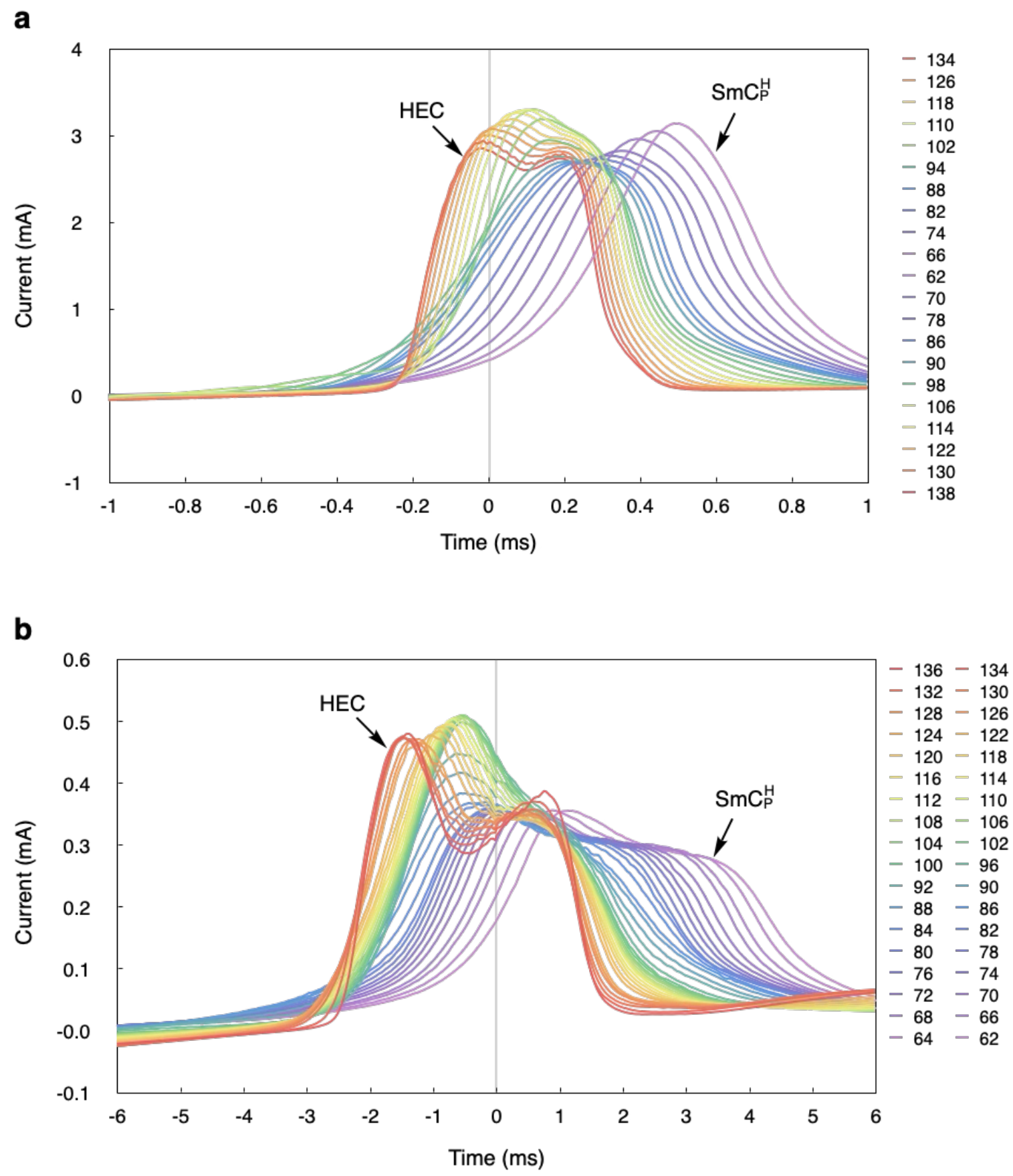

**Figure S17** Comparison of PRC data within the HEC and $SmC_P^H$ ranges for **Compound 1**. PRC data obtained under conditions: (a) $V_{pp}$ = 40 V, $f$ = 100 Hz (b) $V_{pp}$ = 20 V, $f$ = 20 Hz.

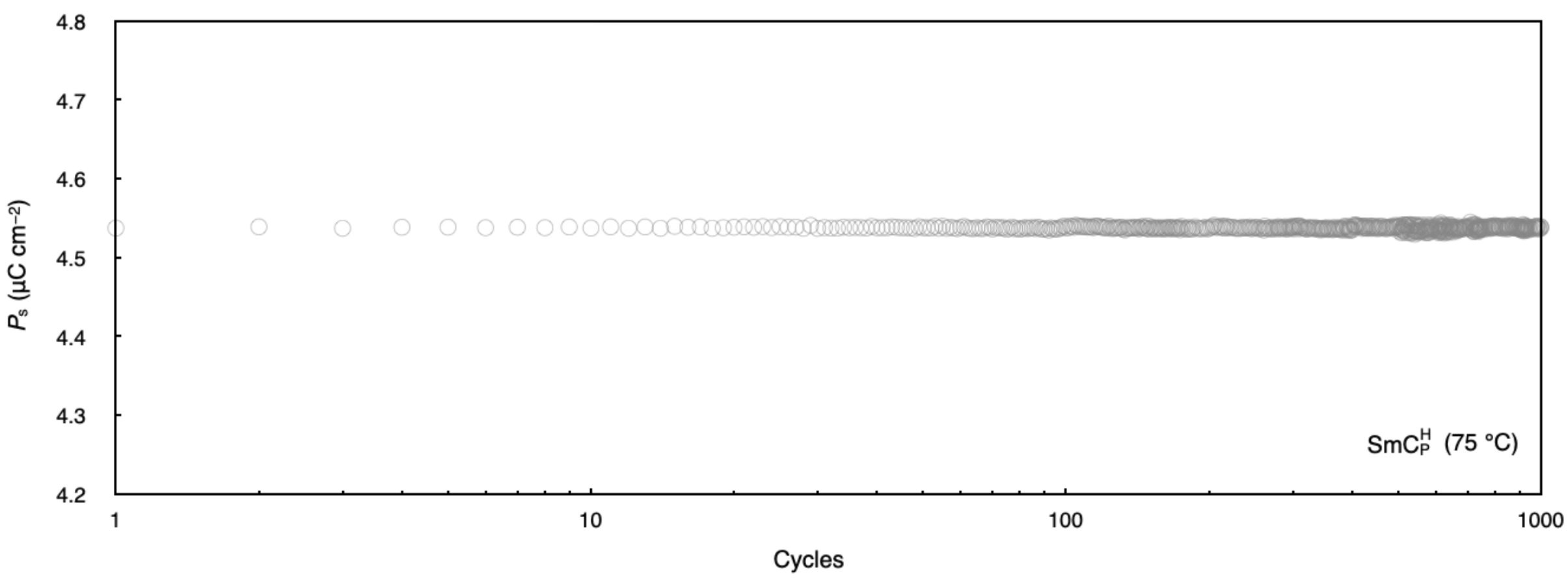

**Figure S18** *E*-field fatigue test in the SmC$_P^H$ (75 °C) for **3-212**.

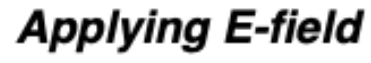

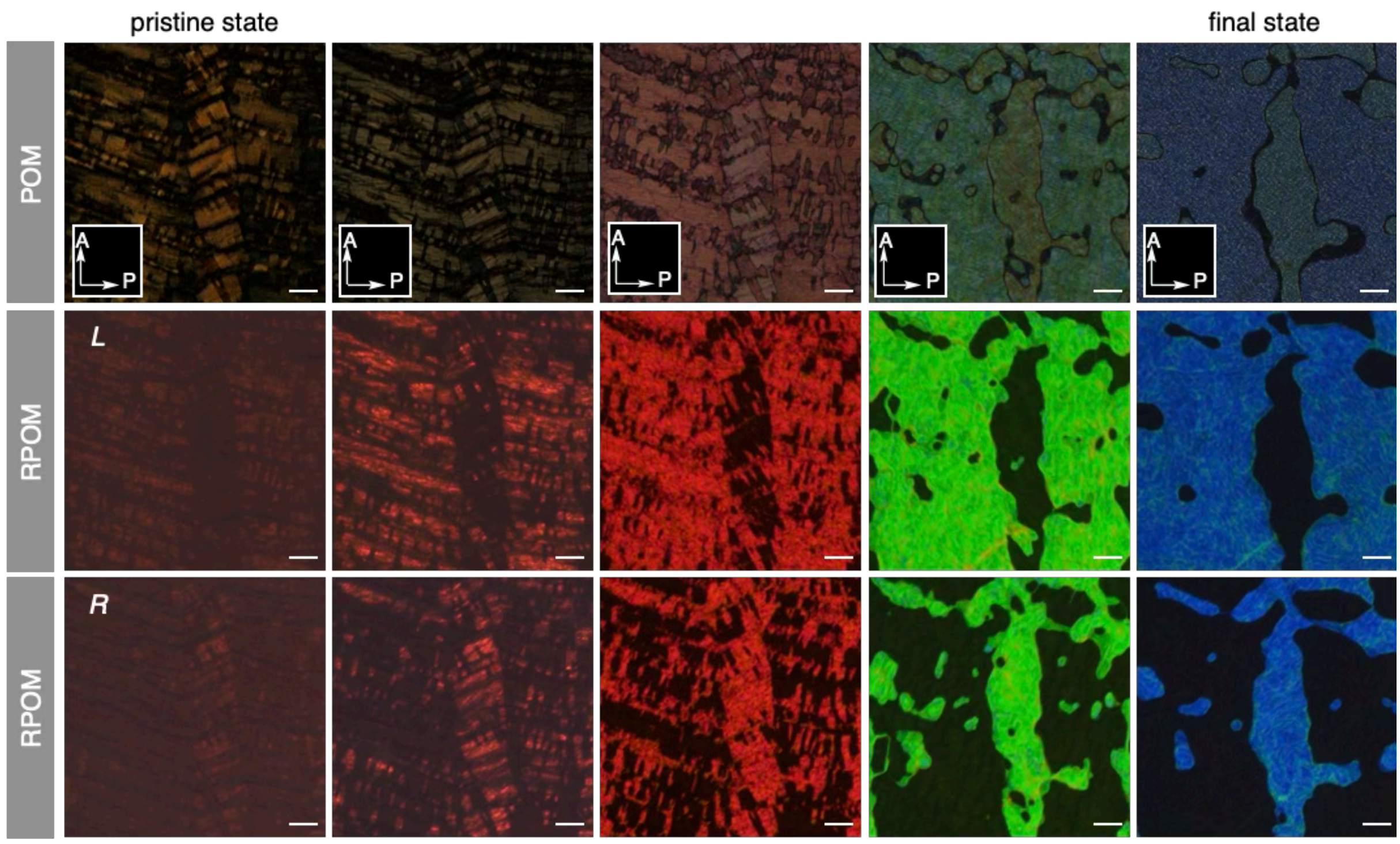

**Figure S19** Changes in POM and RPOM images in the $\mathrm{SmC}_{\mathrm{p}}^{\mathrm{H}}$ phase (75 °C) under AC *E*-field.

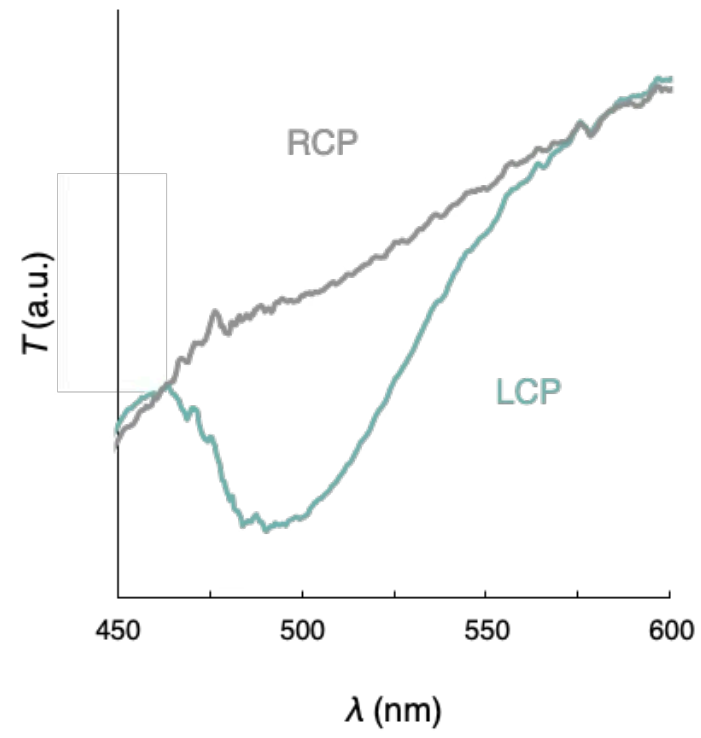

**Figure S20** Transmittance spectra at the local homochiral domain through right- and left-handed circular polarizers (RCP and LCP) in the $SmC_p^H$ phase (75 °C) under *E*-field.

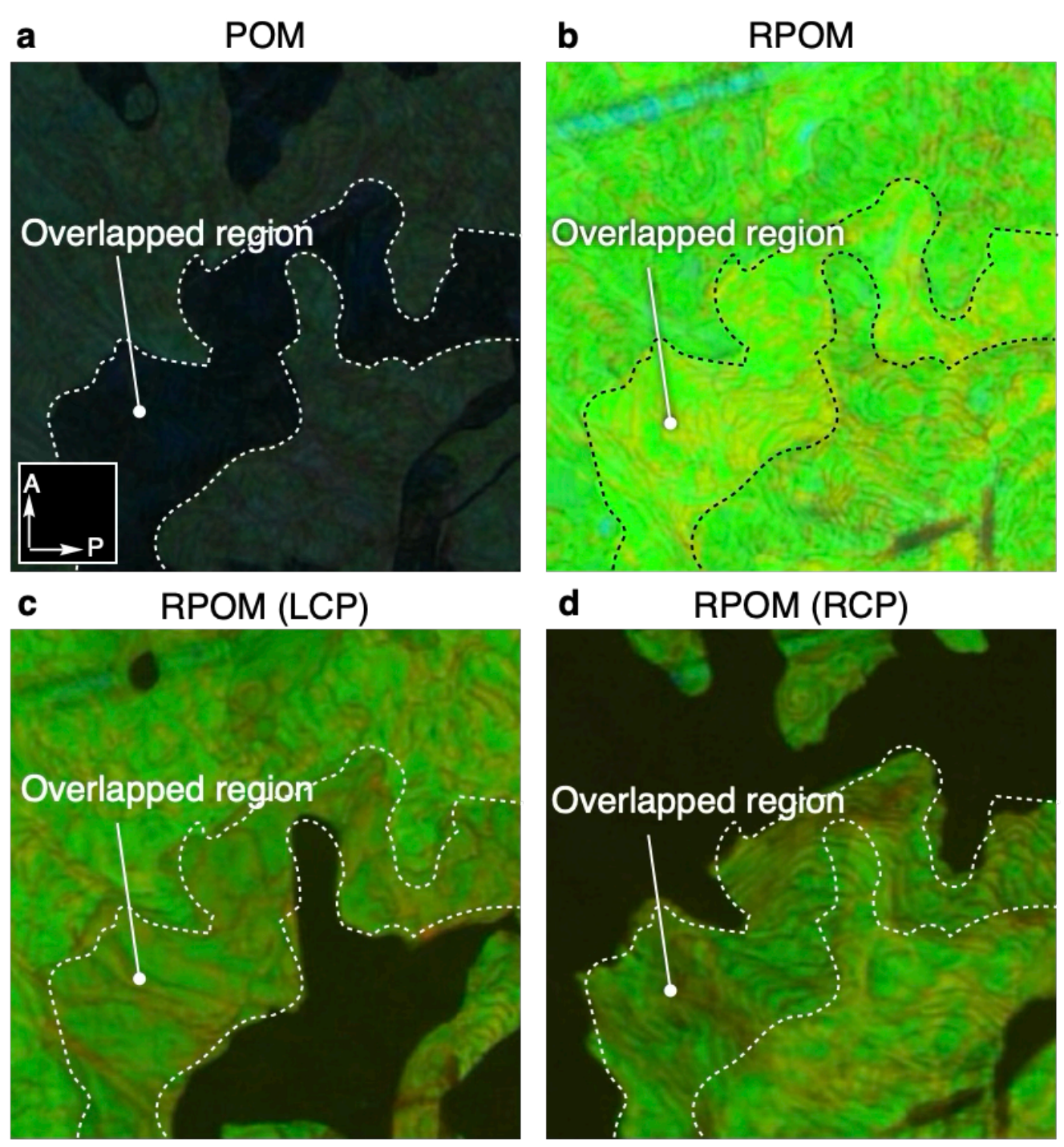

**Figure S21** Extend POM images in the $SmC_P^H$ phase (75 °C) under *E*-field. a) POM, b) reflective mode POM (RPOM) and RPOM images through left- (c) and right- (d) handed circular polarizers.

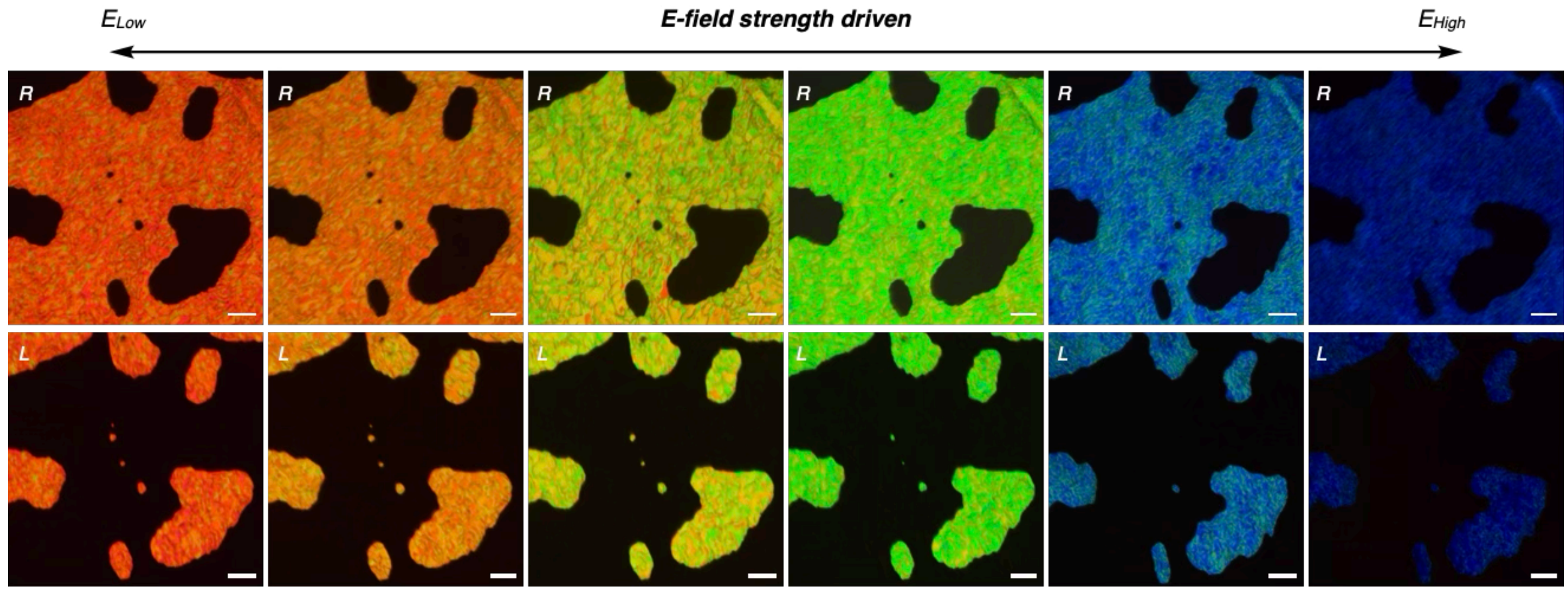

**Figure S22** Extend RPOM images of the $SmC_P^H$ phase (75 °C) under *E*-field. These images were taken in thin LC cell (4.5 μm) through RCP and LCP.

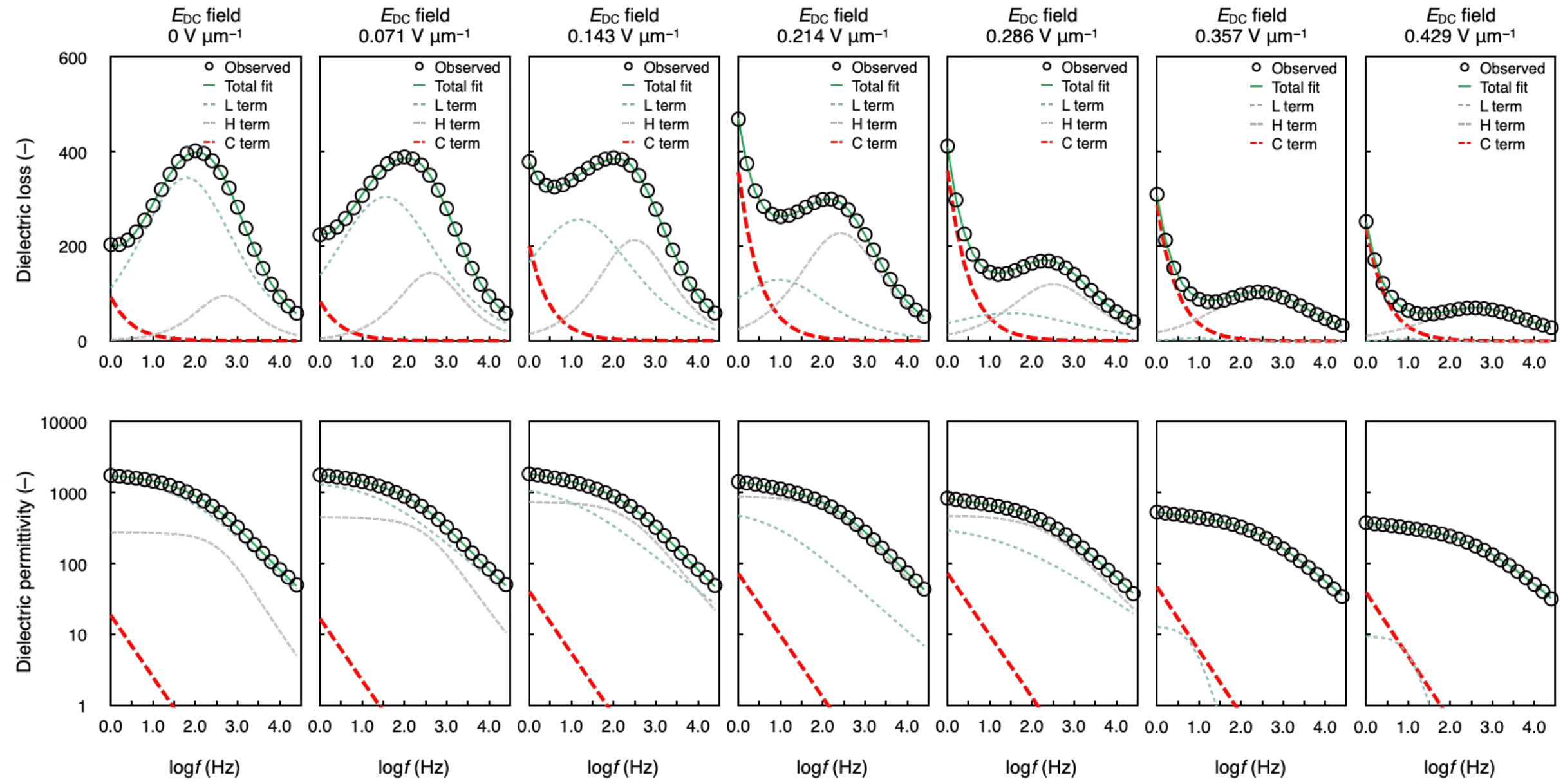

**Figure S23** DC-bias-dependent BDS analysis in the $SmC_P^H$ phase (75 °C) for **Compound 1**. Frequency dependence of the dielectric loss ($\varepsilon''$) (upper panels), and dielectric permittivity ($\varepsilon'$) (lower panels), measured using a small AC probing voltage (100 mV) superimposed on the indicated $E_{DC}$- fields. Open circles represent the experimental data, and solid lines represent the total fitted spectra obtained using two Havriliak–Negami relaxation contributions together with a low-frequency conductivity-related term. The dashed lines show the individual contributions of the low-frequency relaxation mode (L term), high-frequency relaxation mode (H term), and conductivity term (C term).

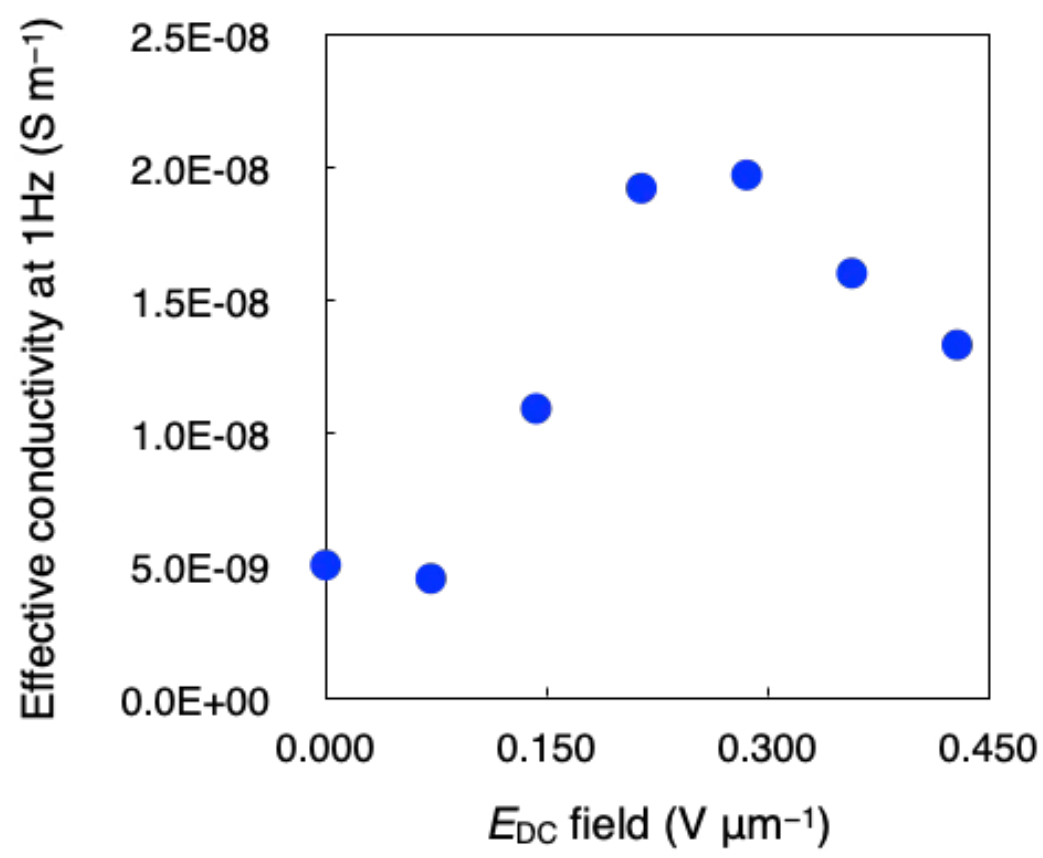

**Figure S24** DC-field dependence of the effective conductivity at 1 Hz, evaluated from the dielectric-loss contribution of the fitted conductivity term (C term).

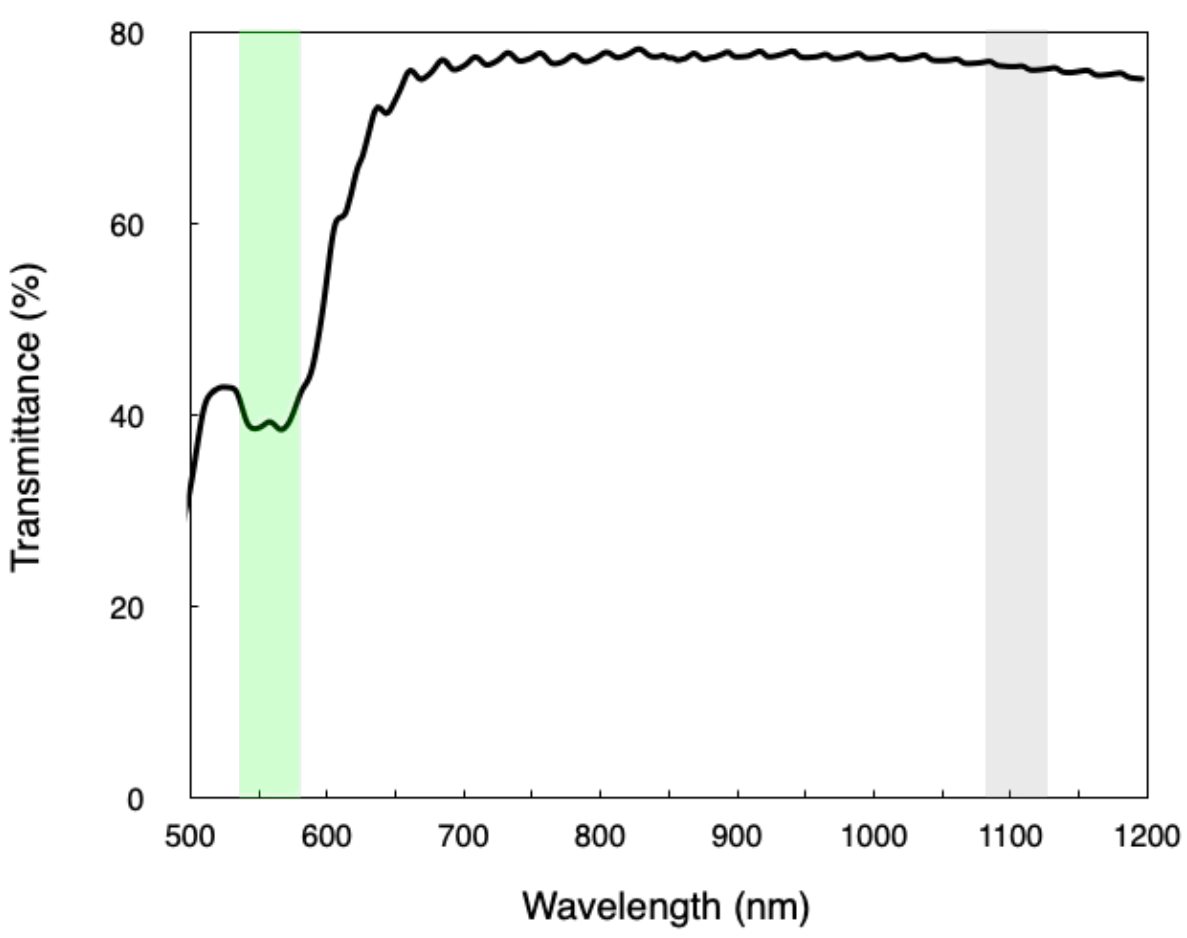

**Figure S25** UV-Vis-NIR spectra recorded in the $\mathrm{SmC_P^H}$ phase (75 °C) for **Compound 1** under *E*-field of 0.75 $V_{pp}$ μm$^{-1}$.

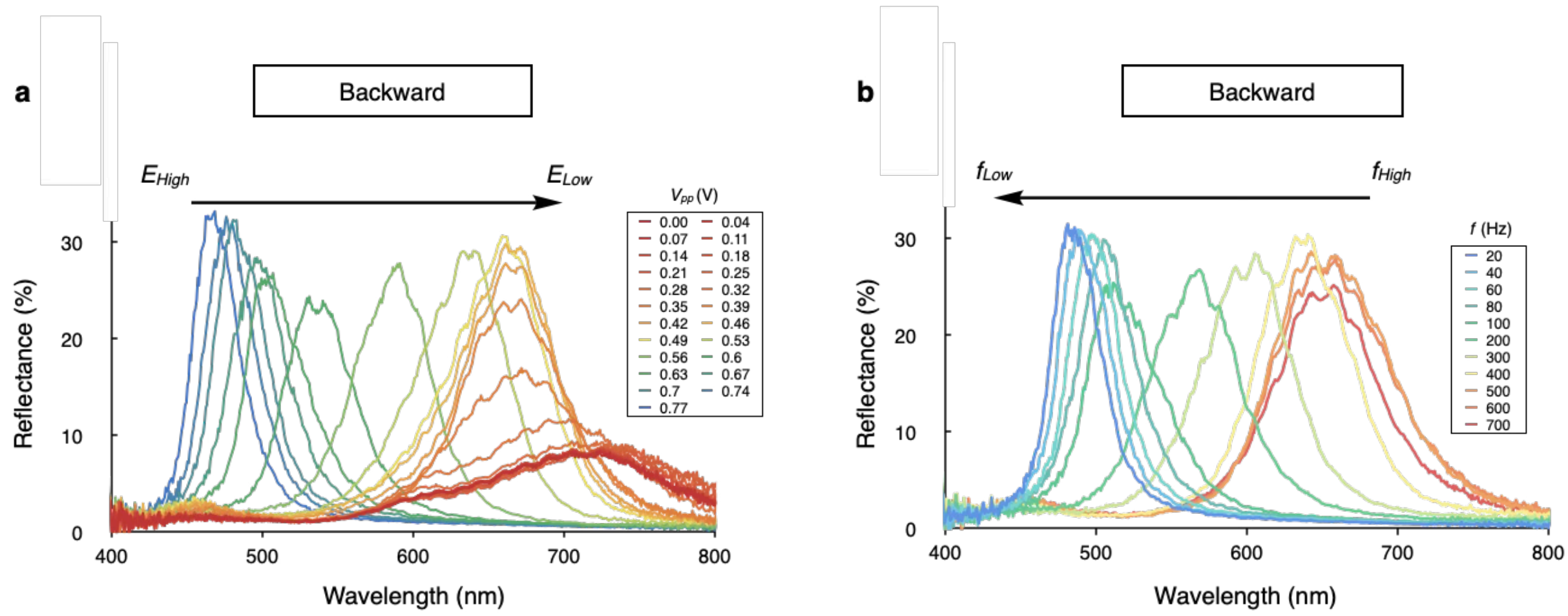

**Figure S26** Backward process of *E*-field- (a) and Frequency- (b) driven modulation for reflection band in the $SmC_P^H$ phase (75 °C): (a) $E_{High}$ to $E_{low}$, (b) $f_{High}$ to $f_{Low}$.

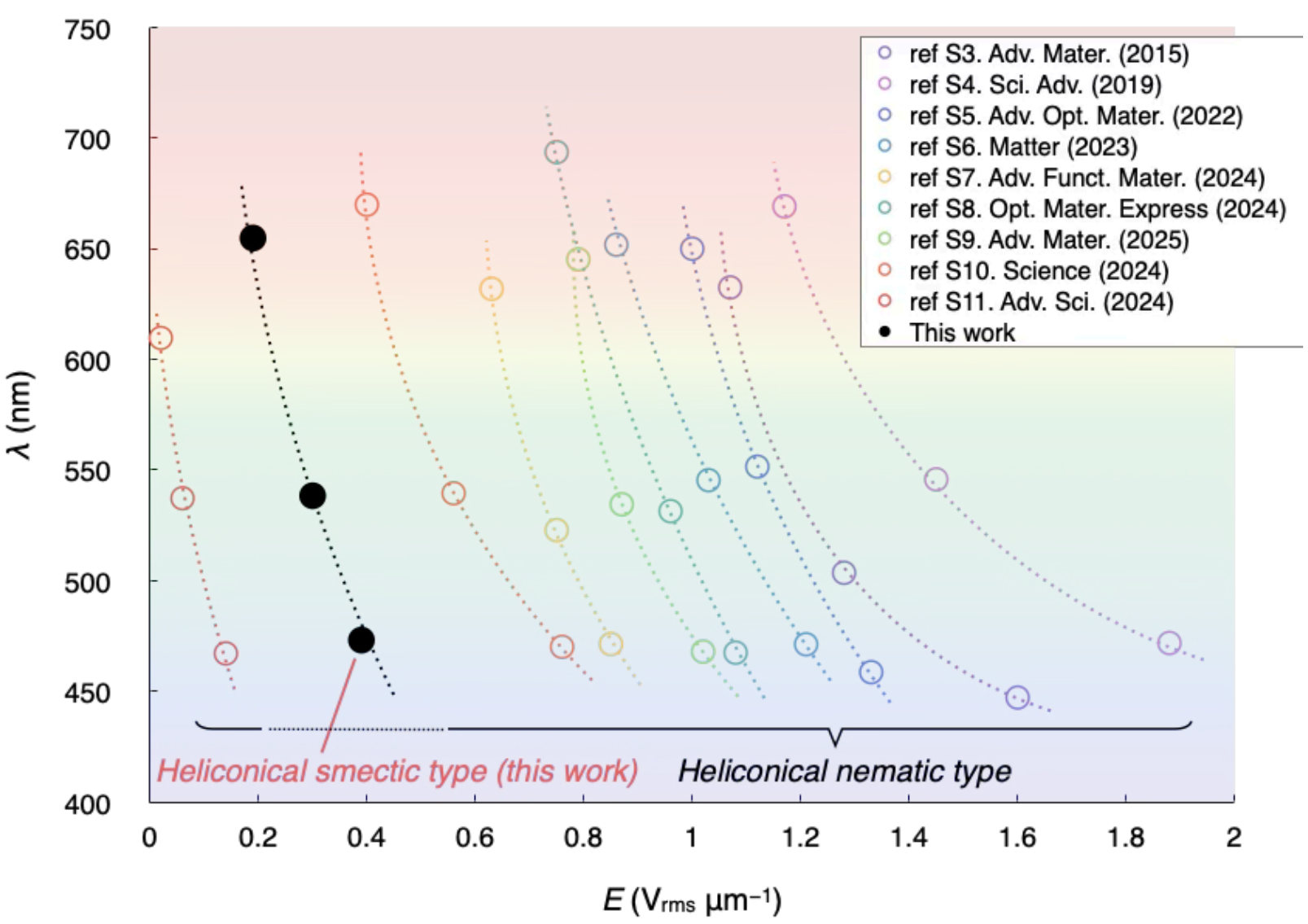

**Figure S27**. *E*-field-dependent peak reflection wavelength ($\lambda$) of representative heliconical nematic-type systems reported in the literature [S3–S11] and the present heliconical smectic-type system. Data were extracted from the corresponding publications. Dotted curves are guides to the eye. In the heliconical smectic system, the plots are not smoothly connected because of the distinct scaling behavior in different regimes.

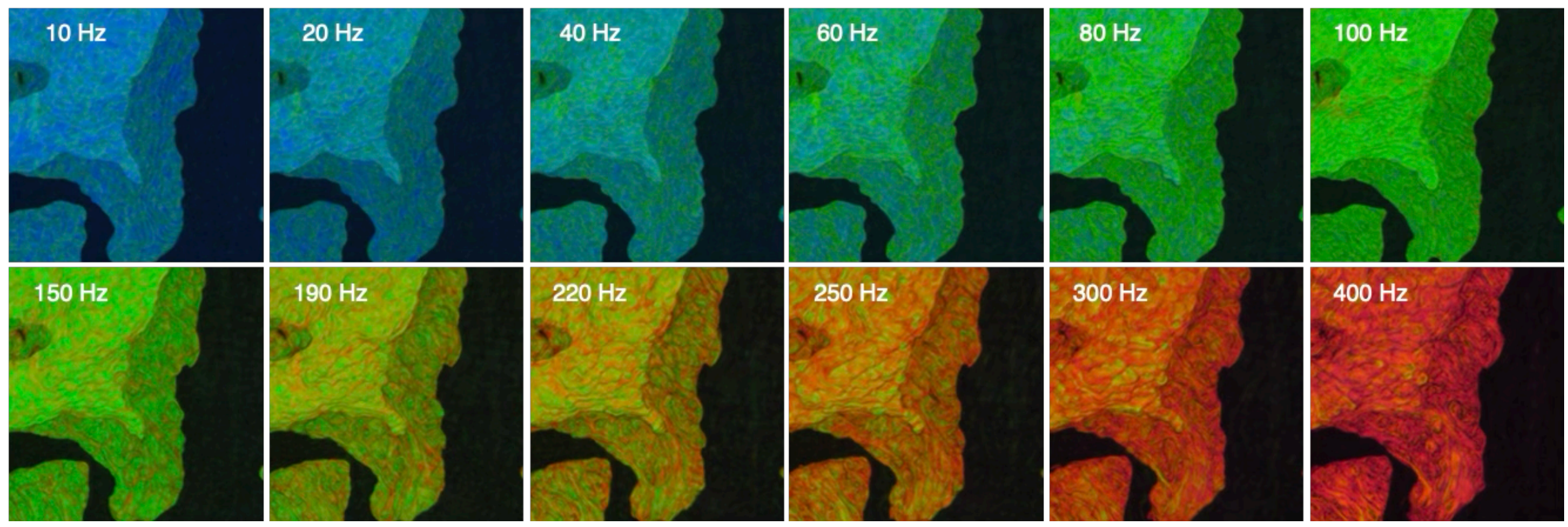

**Figure S28** Frequency-dependent RPOM images ($f$ = 10–400 Hz) taken through the RCP in the $\mathrm{SmC_P^H}$ phase (75 °C) under fixed $E$-field (square waveform, 9.0 $V_{pp}$).

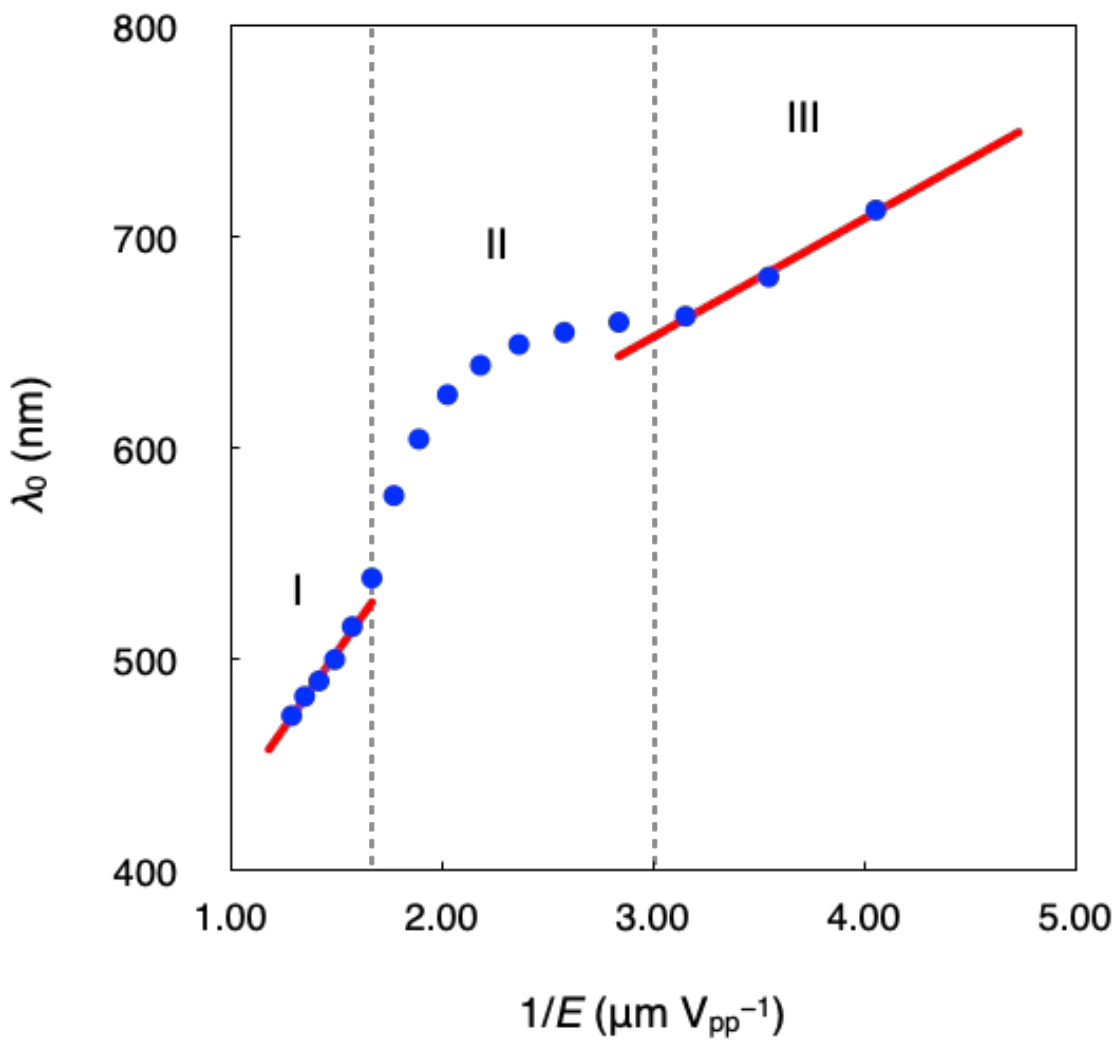

**Figure S29** $\lambda_0$ vs 1/*E* in the $SmC_P^H$ phase (75 °C). The helical pitch exhibits distinct *E*-field-dependent modulations across three regimes (I, II, and III). In regions I and III, $\lambda_0$ exhibits inverse-field scaling ($\lambda_0 \propto 1/E$) and is described by $\lambda_0 = 143/E + 288$ and $\lambda_0 = 56.1/E + 484$, respectively.

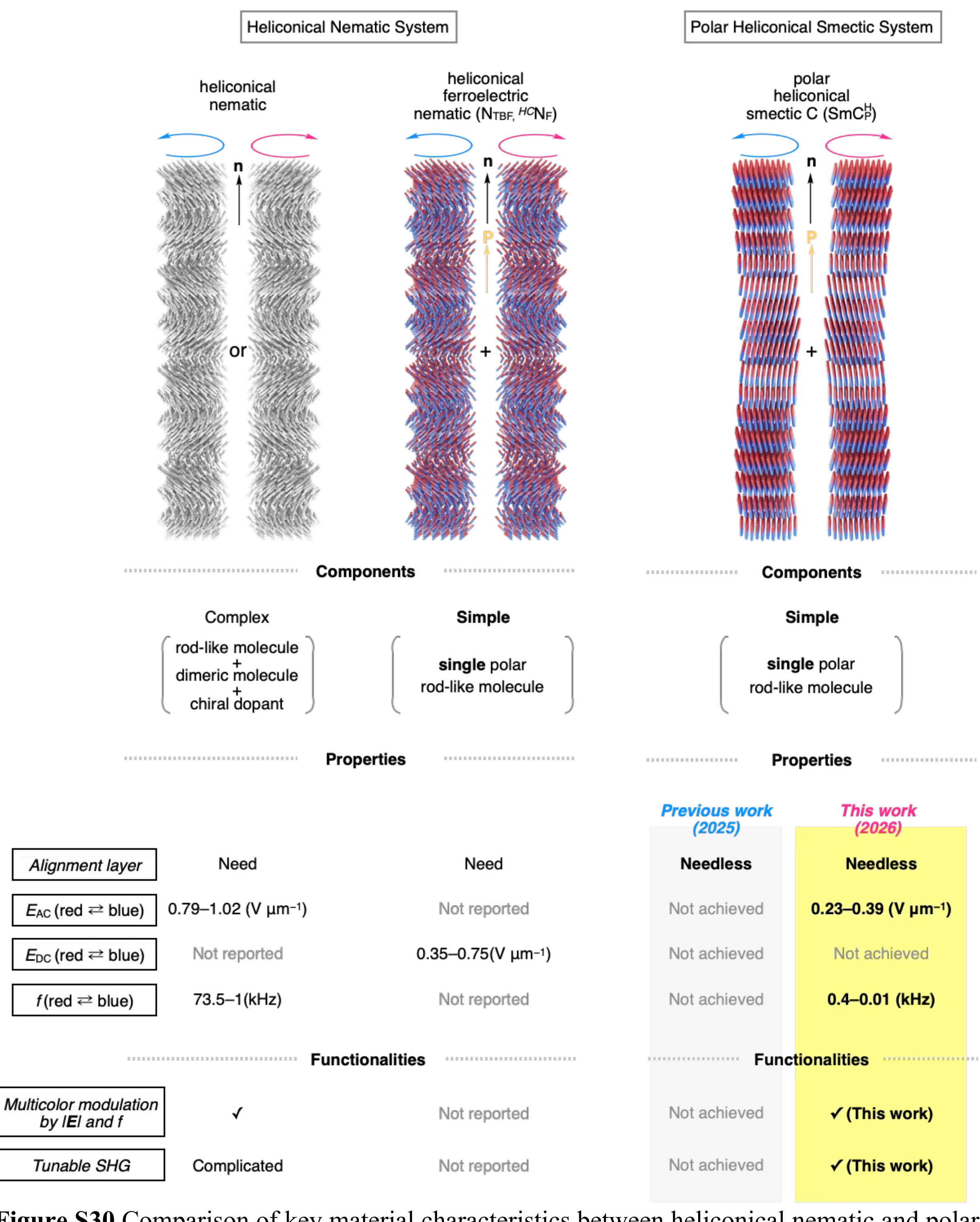

**Figure S30** Comparison of key material characteristics between heliconical nematic and polar heliconical smectic systems.

## Supporting Tables (Table S1, S2)

**Table S1** Phase transition temperature (°C) and enthalpy changes (kJ mol$^{-1}$, in parentheses) for **Compound 1**. Rate: 10 K min$^{-1}$ and 1 K min$^{-1}$.

| Process | K | | $SmC_P^H$ | | HEC | | $N_{TBF}$ | | $SmA_F$ | | $N_x$ | | N | | IL |
|---|---|---|---|---|---|---|---|---|---|---|---|---|---|---|---|
| 1C | • | 58.0 (17.7) | • | 88.8 (0.07) (0.03)[b] | • | 140[a] (i) | • | 143.0[a] (ii) | • | 156.8[a,b,c] (iii) | • | 157.2[a,b,c] (iv) | • | 206 (0.97) | • |
| 2H | • | 84.3 (−22.4) | – | | • | 136.9[a] (v) | • | –[d] (vi) | • | 156.9[a,b,c] (vii) | • | 157.3[a,b,c] (viii) | • | 205 (−0.97) | • |

[a] Overlapping peaks on DSC trace, temperature taken from POM and/or DSC. i + ii = 0.05 kJ mol$^{-1}$, iii + iv = 1.17 kJ mol$^{-1}$, v + vi = −0.05 kJ mol$^{-1}$, vii + viii = −0.145 kJ mol$^{-1}$. [b] Rate: 1 K min$^{-1}$. [c] $T_{peak\text{-}top}$. [d] Undeterminable.

**Table S2** The molecular parameters of energy-minimized conformations calculated by DFT for **Compound 1** and **Model I**.

| Entry | Vector X | Vector Y | Vector Z | $\mu$ (D) | $\beta$ (deg)[a] |
|---|---|---|---|---|---|
| **Compound 1** | 10.0869 | −4.1888 | −1.8005 | 11.07 | 24.32 |
| **Model I**[b] | 9.6458 | −2.8264 | −1.3728 | 10.24 | 18.04 |

[a] An angle between the permanent dipole moment (μ) and long molecular axis; [b] [S12]

**Supporting Videos (Video S1)**

**Video S1** Multicolor modulation in the $SmC_P^H$ phase of **Compound 1** (75 °C). Conditions: ITO glass cell (gap: 14.1 $\mu m^{-1}$), frequency: 100 Hz, *E*-field (square waveform): 0 to 11 $V_{pp}$.

**Video S2** Multicolor modulation in the $SmC_P^H$ phase of **Compound 1** (75 °C). Conditions: ITO glass cell (gap: 14.1 $\mu m^{-1}$), frequency: 100 Hz, *E*-field (square waveform): 11 to 0 $V_{pp}$.

**Video S3** Multicolor modulation in the $SmC_P^H$ phase of **Compound 1** (75 °C). Conditions: ITO glass cell (gap: 14.1 $\mu m^{-1}$), *E*-field (square waveform): 9 $V_{pp}$, frequency: 10 to 400 Hz.

**Video S3** Multicolor modulation in the $SmC_P^H$ phase of **Compound 1** (75 °C). Conditions: ITO glass cell (gap: 14.1 $\mu m^{-1}$), *E*-field (square waveform): 9 $V_{pp}$, frequency: 400 to 10 Hz.

**Video S4** POM texture (reflective mode) recorded thorough a left-handed circular polarizer in the $SmC_P^H$ phase of **Compound 1** (75 °C). Conditions: ITO glass cell (gap: 14.1 $\mu m^{-1}$), *E*-field (square waveform): 11 $V_{pp}$, frequency: 100 Hz.

**Video S5** POM texture (reflective mode) recorded thorough a right-handed circular polarizer in the $SmC_P^H$ phase of **Compound 1** (75 °C). Conditions: ITO glass cell (gap: 14.1 $\mu m^{-1}$), *E*-field (square waveform): 11 $V_{pp}$, frequency: 100 Hz.

**Supporting References**